\documentclass[runningheads]{svmult}

\usepackage{makeidx}   % allows index generation
\usepackage{graphicx}  % standard LaTeX graphics tool
\usepackage{subeqnar}  % subnumbers individual equations
\usepackage{multicol}  % used for the two-column index
\usepackage{physprbb}  % modified textarea for proceedings,
\makeindex             % used for the subject index
\newcommand{\figwidth}{\textwidth}
\newcommand{\medfigwidth}{0.7\textwidth}
\newcommand{\smallfigwidth}{0.5\textwidth}

\begin{document}
\title*{Vortex phases}
\toctitle{Vortex phases}
% allows explicit linebreak for the table of content
%
%
\titlerunning{Vortex phases}
% allows abbreviation of title, if the full title is too long
% to fit in the running head
%
\author{T. Giamarchi\inst{1}
\and S. Bhattacharya\inst{2}
}
\authorrunning{T. Giamarchi and S. Bhattacharya}

\institute{Laboratoire de Physique des Solides, CNRS-UMR8502, UPS B\^at 510, 91405 Orsay, France
\and NEC Research Institute, Princeton, New Jersey 08540, USA}

\maketitle              % typesets the title of the contribution

\begin{abstract}
These lecture notes are meant to provide a pedagogical
introduction, and present the latest theoretical and experimental
developments on the physics of vortices in type II
superconductors.
\end{abstract}

\section{Why and what in vortices}

\begin{figure}
\centerline{\includegraphics[width=\figwidth]{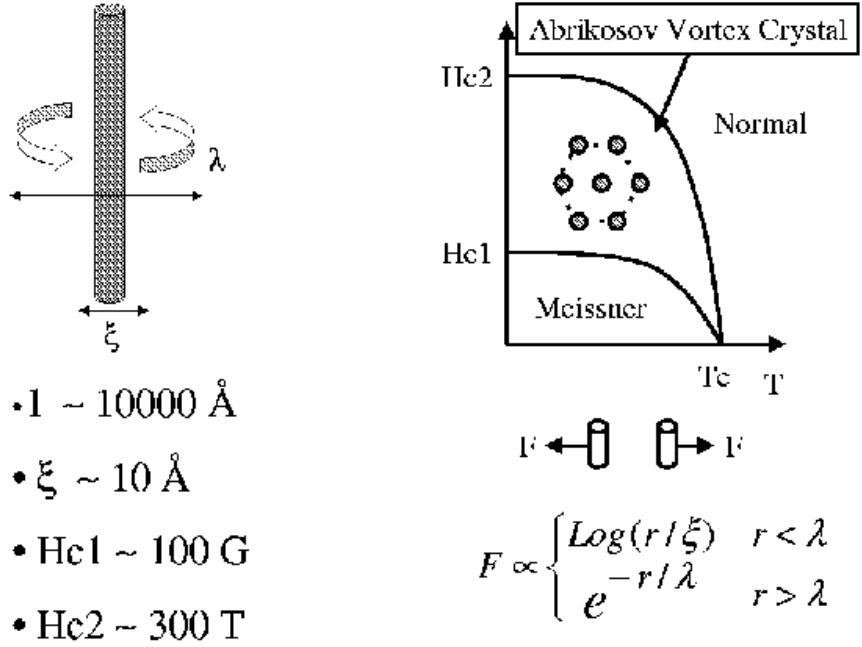}}
\caption{\label{fig:vortex} The structure of a vortex, with a core
size $\xi$ and supercurrent running over a radius $\lambda$ the
penetration length. Typical values for high Tc are given. Due to
supercurrents vortex repel with a force $F$. This leads to the
naive phase diagram where a crystal of vortex exists between
$H_{c1}$ and $H_{c2}$.}
\end{figure}

The discovery of high Tc superconductors has shattered the
comforting sense of understanding that we had of the phase diagram
and physical properties of type II superconductors, and in
particular of the mixed (vortex) phase in such systems. Indeed, it
was well known since Abrikosov \cite{abrikosov_vortex_first} that
above $H_{c1}$ the magnetic field penetrates under the form a
vortex, made of a filament of radius $\xi$ (the coherence length)
surrounded by supercurrent screening the external field running
over a radius $\lambda$ (the penetration length). Because of the
repulsion between vortices due to supercurrents (see
Fig.~\ref{fig:vortex}), the naive idea is that vortex will form a
perfect triangular crystal (the Abrikosov lattice). This has led
to the phase diagram shown in Fig.~\ref{fig:vortex}, that has been
the cornerstone of our understanding of all type II
superconductors for more than three decades
\cite{tinkham_book_superconductors,DeGennes_supra}. However in
high Tc, one could reach much higher temperatures, and it was soon
apparent that some of the physics linked to the existence of the
thermal fluctuations and disorder was overlooked. This led to a
burst of investigations, both theoretical and experimental, to
understand the physical properties of such vortex matter. Of
course, high Tc were not the only field of investigations and low
Tc superconductors were reexamined as well, now that we knew what
to look for in them.

\begin{figure}
\centerline{\includegraphics[width=\smallfigwidth]{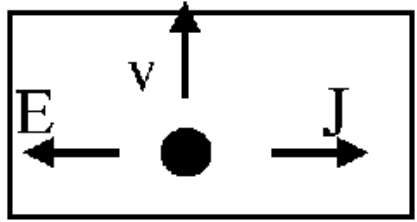}}
\caption{\label{fig:sliding} A current $J$ driven in a
superconductor exert a force on the vortices, making them slide
sideways with a velocity $v$. Since the vortices are flux tubes
this generates an electric field $E$ in the direction of the
current and thus to a finite (and rather bad) resistance.}
\end{figure}

Indeed the vortex phase provides an excellent system for both the
fundamental researcher and one in search of physics with
useful applications. From the fundamental point of view, vortex
matter provides a unique system where one can study a crystal, in
which one can vary the density (the lattice spacing) at the turn
of a knob (simply by varying the magnetic field). In addition,
because this crystal is embedded in a ``space'' with a much
finer lattice constant (the real atomic crystal), it can be
submitted to various perturbations such as disorder, difficult
 to investigate in normal crystals. This provides thus a
unique opportunity to study the combined effect of disorder and
thermal fluctuations on a crystal. From a more practical point of
view, if one has vortices in a superconductor and passes a current
$J$ in them, the current will act as a force on the magnetic tubes
that are the vortices, and they will start to slide. The sliding in
turn generates an electric field $E$, which means that the
superconductor is not superconducting any more, due to the motion
of vortices (see Fig.~\ref{fig:sliding}).
The resistance is rather poor and is simply $\rho = \rho_n
(H/H_{c2})$ (Bardeen-Stephen resistance). In order to get a good
superconductor, it is thus necessary to prevent the vortices from
moving by pinning them. Hence the strong practical incentive to
understand the properties (both static and dynamics) of vortices
in the presence of disorder.

A study of vortices prompts for several questions, that we will
try to address in these notes:
\begin{itemize}
\item What is the effect of disorder on the Abrikosov vortex
crystal ?
\item How to describe the vortex phase ? Does one needs the full
Ginzburg-Landau description or can one use a simplified
description modelling vortices as elastic spaghettis ?
\item What is the phase diagram and the static physical properties
of the vortices in presence of disorder and thermal fluctuations ?
\item What are the dynamical properties ? Is there a linear
response and more generally what is the $I-V$ characteristics ?
\item What is the nature of the vortex system when it is in motion ?
\item Are there links with other physical systems exhibiting the
same competition between crystal order and disorder ?
\end{itemize}

Of course, these questions have been examined intensively in the
last 30 years, and represent an impressive body of research work.
So in these few pages we have to make a choice. Reviews already
exist on vortices
\cite{blatter_vortex_review,brandt_review_superconductors,giamarchi_book_young,%
nattermann_vortex_review} so we will try to present in these notes
the basic ideas enabling the reader to understand the concept
behind the variety of studies, and then bring the reader up to
date with the recent theoretical and experimental developments
that are left out of the previous reviews. The plan of the paper
is as follows: in Section \ref{sec:elastic} we discuss an elastic
description of the vortices, and the issue of lattice melting. In
Section \ref{sec:disorder} we discuss the effects of disorder,
introduce the basic lengthscales and physical concepts and discuss
the limitations of the previously proposed solutions to tackle
this problem. In Section \ref{sec:statics} we discuss the recent
theory of the Bragg glass and compare it with the host of
experimental data on the statics of the vortex lattice. In Section
\ref{sec:dynamics} we discuss the more complicated issue of the
dynamics. Finally conclusions, perspectives and contact with other
physical systems can be found in Section \ref{sec:conclusions}.

\section{Elastic description of vortices} \label{sec:elastic}

A way to get a tractable description of the vortex lattice is to
ignore the microscopic aspects of the superconducting state and
the Ginzburg-Landau description of the vortex, and simply consider
the vortices as an elastic object. The core is like a piece of
string and the supercurrents provide the repulsive (elastic)
forces. Of course such a description is a simplification and
depending on the problem, it will be necessary to check that
important physics has not been left out in the process. However, such a
description has the advantage of being simple enough so that
additional effects such as disorder can be included, and retains
in fact most of the interesting physics for amacroscopic description of the vortex lattice (phase diagram,
imaging, transport). Another advantage is that this
allows us to make contact with a large body of related problems as
will be discussed in section~\ref{sec:conclusions}.

\begin{figure}
\centerline{\includegraphics[width=\figwidth]{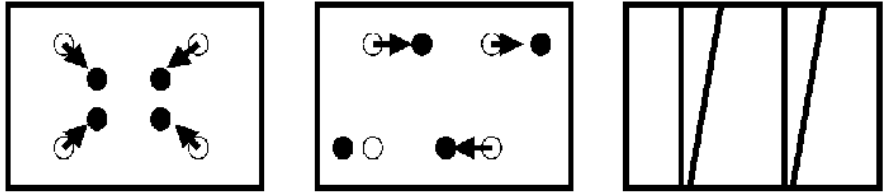}}
\caption{\label{fig:bst} Three different deformations of the
vortex lattice (compression (bulk), shear and tilt) correspond to
three different elastic constants, respectivelly called $c_{11}$,
$c_{66}$ and $c_{44}$.}
\end{figure}

\begin{figure}
\centerline{\includegraphics[width=\smallfigwidth]{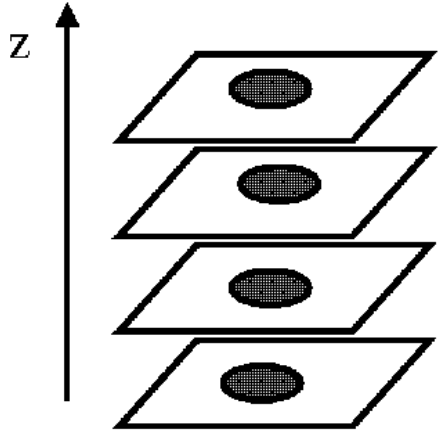}}
\caption{\label{fig:pancakes} In very anisotropic materials, it is
best to see a vortex line as a stack of pancakes vortices coupled
by electromagnetic or Josephson coupling. For most cases however a
line description will be sufficient.}
\end{figure}

We thus describe the vortex system as objects having an
equilibrium position $R_i^0$ (on a triangular lattice for the
vortices, but this is of course general) and a displacement $u_i$
compared to this equilibrium position. $u_i$ is a vector with a
certain number $n$ of components. The vortices being lines $n=2$,
since displacements are on the plane perpendicular to the $z$
axis. The elastic Hamiltonian is
\begin{equation} \label{eq:elas}
H = \frac1{2\Omega} \sum_{\alpha\beta} \sum_q c_{\alpha\beta}(q)
u_\alpha(q)u_\beta(-q)
\end{equation}
where $\alpha=x,y,z$ are the spatial coordinates. The
$c_{\alpha\beta}(q)$ are the elastic constants. The fact that they
have a non trivial dependence on $q$ comes from the long range
nature of the forces between the vortices. The $c$ can be computed
from the microscopic forces between vortices. Standard elasticity
corresponds to $c(q) = c q^2$. Such a behavior will always be
correct at large distance (small $q$) since the forces have a
finite range $\lambda$. In (\ref{eq:elas}) various physical
process have in principle to be distinguished and correspond to
different elastic constants. This corresponds to bulk, shear and
tilt deformations of the vortex lattice as shown on
Fig.~\ref{fig:bst}.
Although these different elastic constants can be widely different
in magnitude (for example bulk compression is usually much more
expensive than shear), this is a simple practical complication
that does not change the quadratic nature of the elastic
Hamiltonian. Such a description is of course also valid for
anisotropic superconductors (such as the layered High Tc ones),
provided that the anisotropy is not too large. If the material is
too layered then it is better to view the vortices as pancakes
living in each plane and coupled by Josephson or electromagnetic
coupling between the planes as shown in Fig.~\ref{fig:pancakes}.
For moderately anisotropic materials viewing the stack of pancakes
as a vortex line is however enough. In this notes we will stick to
this description.

\begin{figure}
\centerline{\includegraphics[width=\figwidth]{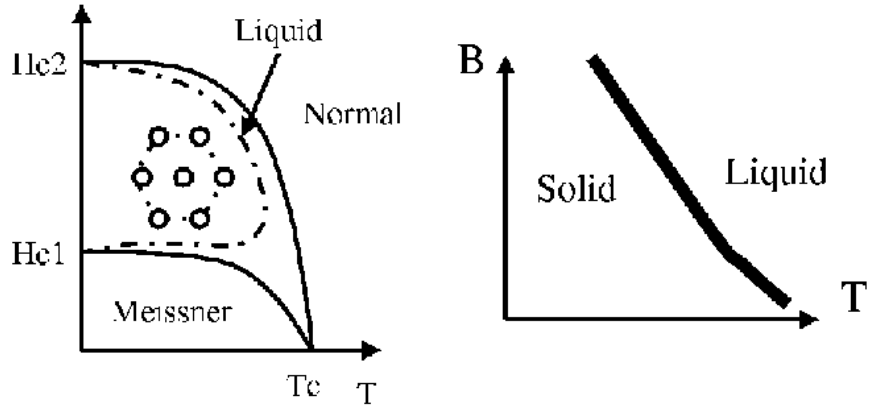}}
\caption{\label{fig:melting} Thermal fluctuations induce melting
of the vortex lattice. On the left the full melting curve is
shown. For high Tc, in practice one is often far from both
$H_{c1}$ and $H_{c2}$ leading to the apparent phase diagram shown
on the right.}
\end{figure}

The melting of the vortex lattice can easily be extracted from the
elastic description. Although a detailed theory of melting is
still lacking, one can use a basic criterion, known as Lindemann
criterion that states that the crystal melts when the thermally
induced displacements a particle in the crystal becomes some
sizeable fraction of the lattice spacing. On a more formal level
the melting is given by
\begin{equation}
\langle u^2 \rangle = l_T^2 = C_L^2 a^2
\end{equation}
which defined the ``effective'' (thermal) size of the particle
(also known as Lindemann length). The proportionality constant
that reproduces correctly the melting is empirically determined to
be $C_L \sim 0.1-0.2$. A simple calculation based on the elastic
the elastic description thus gives
\begin{equation} \label{eq:melteq}
\frac{T_m}{c} = C_L^2 a^2
\end{equation}
showing that the melting temperature goes down as the lattice
spacing goes down (or the magnetic field up). Of course the full
quantitative study for the vortex lattice should be done with the
full fledged elastic Hamiltonian (including bulk, shear, tilt),
but the main conclusions remain unchanged
\cite{houghton_fusion_vortex,nelson_fusion_vortex}. This leads to
the first modification of the naive Abrikosov phase diagram taking
into account the melting shown in Fig.~\ref{fig:melting}.
Close to $H_{c1}$ the elastic constants drop down (since the
vortices get separated by more than $\lambda$ the force between
them is exponentially small), and from (\ref{eq:melteq}) the
crystal also melts, leading the reentrant behavior shown in
Fig.~\ref{fig:melting}.

\begin{figure}
\centerline{\includegraphics[width=\medfigwidth]{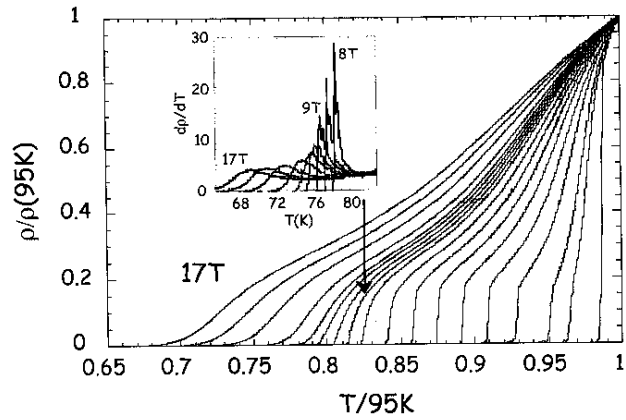}}
\caption{\label{fig:meltexp} T-dependence of ohmic resistance at
various field values. The sudden onset marked by a pronounced knee
marks the melting transition. The inset shows a plot of the
temperature derivative of the resistance which, surprisingly
mimics the specific heat jump across the transition
\cite{kwok_vortex_melting}}
\end{figure}

\begin{figure}
\centerline{\includegraphics[width=\medfigwidth]{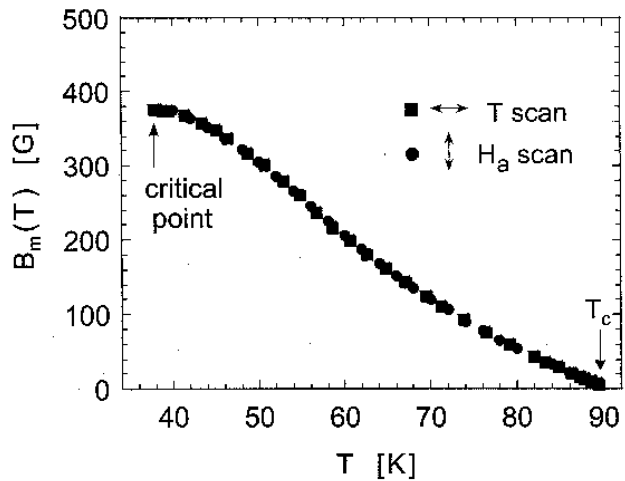}}
\caption{\label{fig:bscco_zeldov} Melting phase diagram obtained
from the jump in local induction for both isothermal and isofield
data \cite{zeldov_diagphas_bisco}. Note the similarity of the
phase boundary with theoretical expectations in
Fig.~\ref{fig:melting} above.}
\end{figure}

\begin{figure}
\centerline{\includegraphics[width=\figwidth]{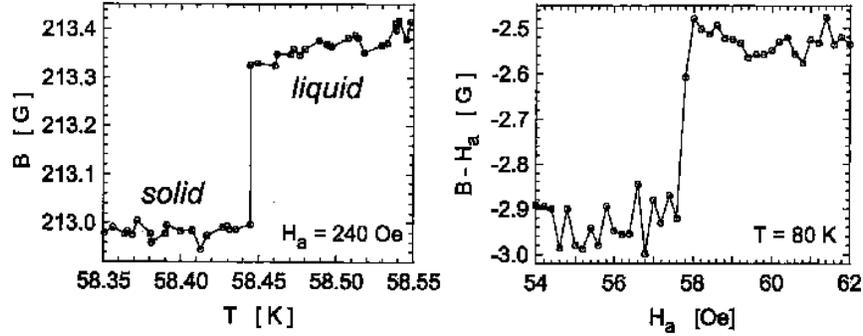}}
\caption{\label{fig:jumps} Typical jumps in local induction for
isothermal and isofield measurements from Hall bar method
\cite{zeldov_diagphas_bisco}. Both the solid and liquid phases
have no measurable pinning in the bulk and the data are
reversible.}
\end{figure}

Early experimental studies
\cite{charalambous_melting_rc,safar_tricritical_prl,safar_transport_tricritical,kwok_vortex_melting}
of the melting transition were based on transport measurements.
The sudden onset of an ohmic resistance was argued to signify
melting and its location in the (H,T) space is the locus of the
melting phase boundary. Typical experimental results are shown in
Fig.~\ref{fig:meltexp}, where the onset is characterized by a
pronounced knee in the resistance.
With increasing external field, the onset moves to lower
temperatures and eventually broadens considerably. This implies
that at sufficiently large fields, the sharp first order
transition crosses over to a more continuous second order like
transition, an effect expected to be the result of disorder (see
later). At lower fields and higher temperatures, the melting
transition is closely approximated by the disorder-free case,
discussed above. Later experiments performed on thermodynamic
quantities such as the magnetization and specific heat confirmed
the first order nature of the melting transition, at least for
weak disorder. Fig.~\ref{fig:bscco_zeldov} shows an experimental
phase diagram of BSCCO obtained by local magnetization using a
novel hall-bar technique \cite{zeldov_diagphas_bisco}.
Figure~\ref{fig:jumps}
shows typical measurements of the jump in the local induction
across this transition, by either changing temperature at a fixed
field or changing field at a fixed temperature.
In both cases, a
positive jump in $B$ is observed in going from the solid to the
liquid phase. Using the Clapeyron- equation :
\begin{equation}
\frac{dH_m}{dT} = -\frac{\Delta S}{\Delta M}
\end{equation}
where $\Delta S$ is the entropy change and $\Delta M$ is the
change in magnetization (density of vortices). The experimental
phase diagram is consistent with theoretical expectations as shown
in Fig.~\ref{fig:melting} above. The negative slope of the melting
curve is consistent with an increase of density in the liquid
phase, curiously akin to a ``water-like'' melting phenomenon. Very
little experimental work is available on the low field reentrant
branch of the melting phase boundary. At low fields the
intervortex interaction is weak and the effects of disorder
dominate. Reentrant phenomena in peak effect (see
Section~\ref{sec:peakeffect}) have been somewhat widely observed
and is thought to be dominated by effects of disorder, rather than
thermal fluctuations.

\begin{figure}
\centerline{\includegraphics[width=\smallfigwidth]{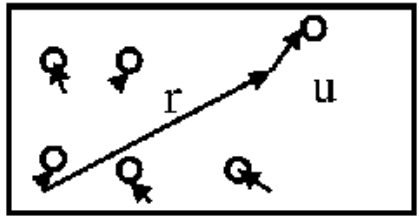}}
\caption{\label{fig:relat} The relative displacement correlation
function $B(r)$ measuring the displacements between two points
separated by a distance $r$ is directly measured in experiments
such as decoration experiments.}
\end{figure}

Finally, besides the phase diagram, what are the physical quantities that one can in principle
compute and that are directly connected to experiments ?
The first important information is the relative displacements correlation function
\begin{equation} \label{eq:relat}
B(r) = \overline{\langle [u(r) - u(0)]^2 \rangle}
\end{equation}
where $\langle \rangle$ is the average over thermal fluctuations
and $\overline{\cdots}$ is the average over disorder (if need be).
(\ref{eq:relat}) indicates how the displacements between two
points in the system separated by a distance $r$ grow (see
Fig.~\ref{fig:relat}).
Decoration experiments provide a direct measure of this
correlation function as we will see. In a perfect crystal $B(r)
=0$, whereas both thermal fluctuations and disorder will make the
displacements grow. How $B(r)$ grows tells us whether the system
is well ordered or not. In a good crystal $B(r)$ will saturate to
a finite value whereas it will grow unboundedly if the perfect
positional order of the crystal is destroyed.

\begin{figure}
\centerline{\includegraphics[width=\smallfigwidth]{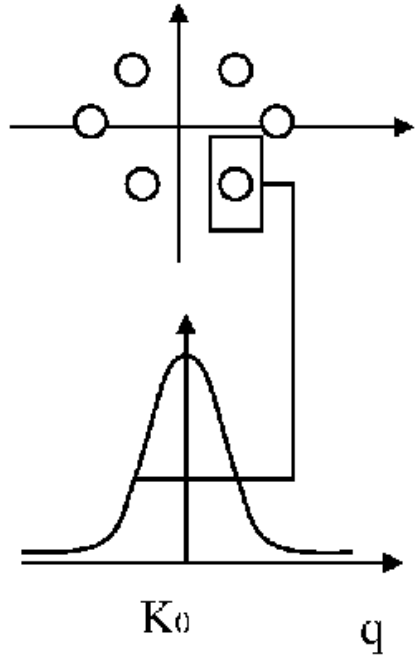}}
\caption{\label{fig:struct} The structure factor measured in
diffraction experiments such as neutrons or X-rays. The shape of a
peak is given by the Fourier transform of the positional
correlation function $C(r)$ (see text).}
\end{figure}

Another important quantity, directly measures in diffraction
experiments is the structure factor
\begin{equation}
S(q) = \overline{\langle \rho(q) \rho(-q) \rangle}
\end{equation}
In a perfect crystal this consists of Bragg peaks at the vectors
$K$ of the reciprocal lattice of the crystal. If one considers one
such peak its shape (as shown on Fig.~\ref{fig:struct})
is the Fourier transform of the positional correlation function
\begin{equation}
C(r) = \overline{\langle e^{i K u(r)} e^{- i K u(0)} \rangle}
\end{equation}
Thus in a perfect crystal $C(r) = 1$ and the Fourier transform is
a $\delta(q)$ Bragg peak. If there are only thermal fluctuations
$C(r\to \infty) \to {\rm Cste}$ (in fact $C(\infty) = e^{-K^2
l_T^2/2}$). The Fourier transform is still a $\delta$ peak but
with a reduced weight which is simply the Debye Waller factor. The
faster $C(r)$ decreases the more disordered is the crystal. If
$C(r)$ decreases exponentially to zero with a characteristics
lengthscale $R_a$, the peak in the structure factor is some
patatoidal (lorentzian like) shape with a width $R_a^{-1}$
indicating that the perfect translational order is lost. Although
it is not always true (it is only true for gaussian fluctuations
such as thermal fluctuations) a rule of thumb is
\begin{equation}
C(r) \sim e^{- K^2 B(r) /2}
\end{equation}
showing quite logically that the faster the relative displacements
grow the more positional order (measured by the Bragg peaks) is
destroyed in the system.

\section{Disorder, basic lengths and open questions}\label{sec:disorder}

The next task is to consider the effects of disorder. In real
systems, disorder exists in all varieties in the underlying atomic
crystal: vacancies, interstitials, lattice dislocations, grain
boundaries, twin boundaries, second-phase precipitates, etc. In
high quality single crystals it is possible to limit dominant
disorder to point like impurities. Additionally, point like
disorder can be, and sometimes is, intentionally added to systems
in the form of electron irradiation in the case of the cuprates
and/or substitutional atomic impurities of various kinds in low Tc
systems. More artificial disorder can be
introduced for example by heavy ions irradiation that produce
columns of defects in the material. We will briefly mention the
consequences of such artificial disorder in Section
\ref{sec:conclusions}, but most of these notes will be devoted to the
effects of point like impurities.

\begin{figure}
\centerline{\includegraphics[width=\smallfigwidth]{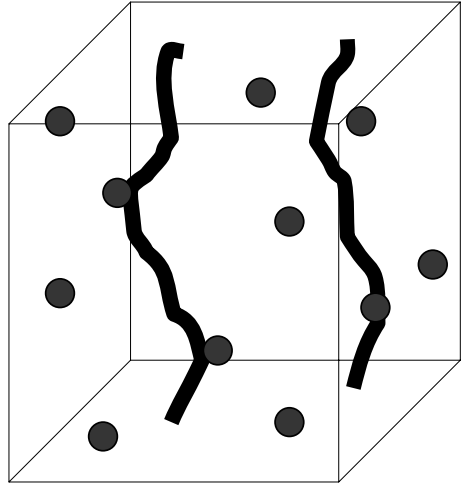}}
\caption{\label{fig:impur} Point like impurities (small circles)
act as pinning centers for the vortices.}
\end{figure}

Such impurities (as shown on Fig.~\ref{fig:impur})
can be modeled by a random potential $V(r)$ coupled directly to
the density $\rho(r) = \sum_i \delta(r -R^0_i - u_i)$ of vortex
lines
\begin{equation} \label{eq:dis}
H_{dis} = \int d^d r V(r) \rho(r)
\end{equation}
In principle one has ``just'' to add (\ref{eq:dis}) to
(\ref{eq:elas}) and solve. Unfortunately the coupling of the
displacements to disorder is highly non linear (since it occurs
inside a $\delta$ function), and thus this is an horribly
complicated problem. Physically this traduces the fact that there
is a competition between the elastic forces that want the system
perfectly ordered and the disorder that let the lines meander.
This competition is bound to lead to complicated states where the
system tries to compromise between these two opposite tendencies.

In order to understand the basic physics of such problem a simple
(but ground breaking !) scaling argument was put forward by Larkin
\cite{larkin_70}. To know whether the disorder is relevant and
destroys the perfect crystalline order, let us assume that there
exists a characteristic lengthscale $R_a$ for which the relative
displacements are of the order of the lattice spacing $u(R_a) -
u(0) \sim a$. If the displacements vary of order $a$ over the
lengthscale $R_a$ the cost in elastic energy from (\ref{eq:elas})
is
\begin{equation} \label{eq:loss}
\frac{c}2 R_a^{d-2} a^2
\end{equation}
by simple scaling analysis. Thus in the absence of disorder
minimizing the energy would lead to $R_a =\infty$ and thus to a
perfect crystal. In presence of the disorder the fact that
displacements can adjust to take advantage of the pinning center
on a volume of size $R_a^d$ allow to gain some energy. Since $V$
is random the energy gained by adapting to the random potential is
the square root of the potential over the volume $R_a^d$, thus one
gains an energy from (\ref{eq:dis})
\begin{equation} \label{eq:gain}
- V R_a^{d/2} \rho_0
\end{equation}
Thus minimizing (\ref{eq:loss}) plus (\ref{eq:gain}) shows that
below four dimensions the disorder is always relevant and leads to
a finite lengthscale
\begin{equation}
R_a \sim a \left(\frac{c^2 a^d}{V^2 \rho_0^2}\right)^{1/(4-d)}
\end{equation}
at which the displacements are of order $a$. The conclusion is
thus that even an arbitrarily weak disorder destroys the perfect
positional order below four dimensions, and thus no disordered
crystal can exist for $d \leq 4$. This is an astonishing result,
which has been rediscovered in other context (for charge density
waves $R_a$ is known as Fukuyama-Lee \cite{fukuyama_pinning}
length and for random field Ising model this is the Imry-Ma length
\cite{imry_ma}). Of course it immediately prompt the question of
what is the resulting phase of elastic system plus disorder ?

\begin{figure}
\centerline{\includegraphics[width=\medfigwidth]{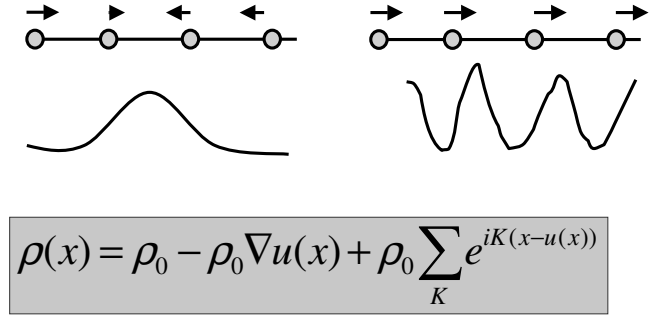}}
\caption{\label{fig:decdens} Various harmonics of the density. If
one is only interesting in variations of the density at
lengthscales large compared to the lattice spacing one has the
standard ``elastic'' expression of the density in terms of the
displacements. In the vortex system however on has to consider
variations of density at lengthscales smaller than the lattice
spacing and higher harmonics are needed
\cite{giamarchi_vortex_short,giamarchi_vortex_long}.}
\end{figure}

Since solving the full problem is tough another important step was
made by Larkin \cite{larkin_70,larkin_ovchinnikov_pinning}. For
small displacements he realized that (\ref{eq:dis}) could be
expanded in powers of $u$ leading to the simpler disorder term
\begin{equation}
H_{\rm larkin} = \int d^dr f(r) u(r)
\end{equation}
where $f(r)$ is some random force acting on the vortices. Because
the coupling to disorder is now linear in the displacements the
Larkin Hamiltonian is exactly solvable. Taking a local random
force $\overline{f(r)f(r')} = \Delta \delta(r-r')$ gives for the
relative displacements correlation function and structure factor
\begin{eqnarray}
B(r) &=& B_{\rm thermal}(r) + \frac{\Delta}{c^2} r^{4-d} \\
C(r) &=& e^{-K^2 B(r) /2} \simeq e^{- r^{4-d}}
\end{eqnarray}
where $B_{\rm thermal}$ are the displacements in the absence of
disorder due to thermal fluctuations (which remain bounded in
$d>2$ and are thus negligeable at large distance compared to the
disorder term). Thus the solution of the Larkin model confirms the
scaling analysis: (i) displacements do grow unboundedly (as a
power law) and thus perfect positional of the crystal is lost;
(ii) the lengthscale at which the displacements are of the order
of the lattice is the similar to the one given by the scaling
analysis. In addition the Larkin model tells us how fast the
positional order is destroyed: the displacements grow as power law
thus the positional is destroyed exponentially fast, leading to
peaks in the structure factor of width $R_a^{-1}$. However all
these conclusions should be taken with a grain of salt. Indeed the
Larkin model is an expansion in powers of $u$, and thus cannot be
valid at large distance (since the displacements grow unboundedly
the expansion has to break down at some lengthscale). What is this
characteristic lengthscale ? A naive expectation is that the
Larkin model cease to be valid when the displacements are of order
$a$ i.e. at $r=R_a$. In fact this is too naive as was noticed by
Larkin and Ovchinikov. To understand why, in a transparent way,
let us rewrite the density of vortices in a way more transparent
than the original form
\begin{equation} \label{eq:densbas}
\rho(x) = \sum_i \delta(r - R_i^0 - u_i)
\end{equation}
Taking the continuum limit for the displacements in
(\ref{eq:densbas}) should be done with care since one is
interested in variations of the density at scales that can be {\it
smaller} than the lattice spacing. A very useful way to rewrite
the density is \cite{giamarchi_vortex_short,giamarchi_vortex_long}
(see also \cite{haldane_bosons,nattermann_pinning}):
\begin{equation} \label{eq:densexp}
\rho(r) = \rho_0 - \rho_0\nabla\cdot u(r) + \rho_0 \sum_K e^{i
K(r-u(r))}
\end{equation}
which is a decomposition of the density in Fourier harmonics
determined by the periodicity of the underlying perfect crystal as
shown on Fig.~\ref{fig:decdens}.
The sum over the reciprocal lattice vectors $K$ obviously
reproduces the $\delta$ function peaks of the density
(\ref{eq:densbas}). If one considers particles with a given size
(for vortices it is the core size $\xi$) then the maximum $K$
vector in the sum should be
\begin{equation}
K_{max} \sim 2\pi/\xi
\end{equation}
in order to reproduce the broadening of order $\xi$ of the peaks
in density. This immediately allows us to reproduce the Larkin
model by expanding (\ref{eq:dis}) using (\ref{eq:densexp})
\begin{equation}
\rho_0 \int d^d r \sum_K e^{i K(r - u(r))} V(r)
\end{equation}
in powers of $u$. Clearly the expansion is valid as long as
$K_{max} u \ll 1$ This will thus be valid up to a lengthscale
$R_c$ such that $u(R_c)$ is of the order of the {\it size of the
particles} $\xi$. Note that this lengthscale is different (and
quite generally smaller) than the lengthscale $R_a$ at which the
displacements are of the order of the lattice spacing. The Larkin
model cease to be valid way before the displacements become of the
order of $a$ and thus {\it cannot be used} to deduce the behavior
of the positional order at large length scale. In addition it is
easy to check that because the coupling to disorder is linear in
the Larkin model, this model does not exhibit any pinning. Any
addition to an external force leads to a sliding of the vortex
lattice. It thus seems that this model is not containing the basic
physics needed to describe the vortex lattice. In a masterstroke
of physical intuition Larkin realized that the lengthscale at
which this model breaks down is precisely the lengthscale at which
pinning appears \cite{larkin_ovchinnikov_pinning}. The lengthscale
$R_c$ is thus the lengthscale above which various chunks of the
vortex system are collectively pinned by the disorder. A simple
scaling analysis on the energy gained when putting an external
force
\begin{equation}
H = \int d^dr F_{\rm ext} u(r)
\end{equation}
allows to determine the critical force needed to unpin the
lattice. Assuming that the critical force needed to unpin the
lattice is when the energy gained by moving due to the external
force is equal to the balance of elastic energy and disorder
$\frac{c}2 \xi^2 R_c^{d-2}$, one obtains
\begin{equation}
J_c \propto \frac{c \xi}{R_c^2}
\end{equation}
This is the famous Larkin-Ovchinnikov relation which allows to
relate a dynamical quantity (the critical current at $T=0$ needed
to unpin the lattice) to purely static lengthscales, here the
Larkin-Ovchinikov length at which the displacements are of the
order of the size of the particle. Let us insist again that this
lengthscale controling pinning is quite different from the one
$R_a$ at which displacements are of the order of the lattice
spacing at that controls the properties of the positional order.

\begin{figure}
\centerline{\includegraphics[width=\medfigwidth]{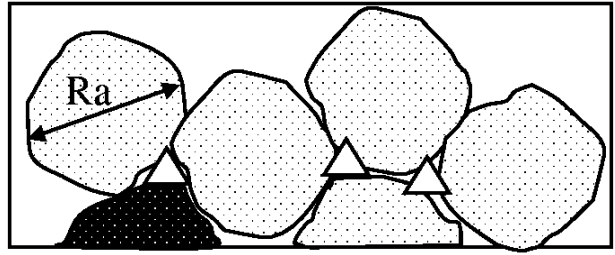}}
\caption{\label{fig:crystallite} The (incorrect) physical image
that was the commonly accepted view of what a disordered elastic
system would look like. The crystal would be broken into
crystallites of size $R_a$ by the disorder. Dislocations would be
generated by the disorder at the same lengthscale.}
\end{figure}

The lack of efficiency of the Larkin model to describe the
behavior of displacements beyond $R_c$ still leaves us with the
question of the nature of the positional order at large distances.
However, extrapolating naively the Larkin model would give a power
law growth of displacements. Such behavior is in agreement with
exact solutions of interface problems in random environments (so
called random manifold problems) and solutions in one spatial
dimension. It was thus quite naturally assumed that an algebraic
growth of displacements was the correct physical solution of the
problem, and thus that the positional order would be destroyed
exponentially beyond the length $R_a$. This led to an image of the
disordered vortex lattice that consisted of a crystal ``broken''
into crystallites of size $R_a$ due to disorder. To reinforce this
image (incorrect) ``proofs'' were given
\cite{fisher_vortexglass_long} to show that due to disorder
dislocations would be generated at the lengthscale $R_a$ (even at
$T=0$) further breaking the crystal apart and leaving no hope of
keeping positional order beyond $R_a$. A summary of this
(incorrect) physical image is shown on Fig.~\ref{fig:crystallite}.

\begin{figure}
\centerline{\includegraphics[width=\figwidth]{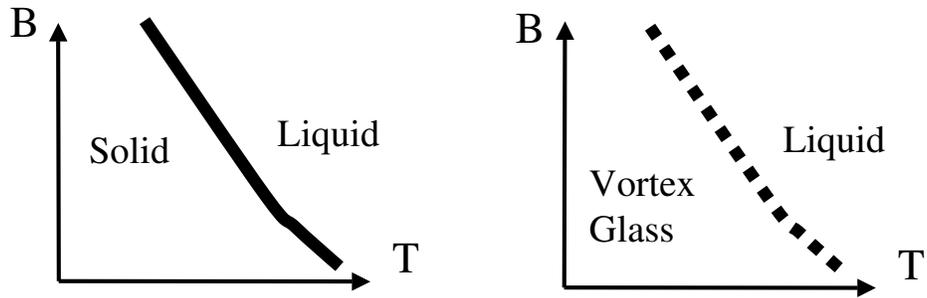}}
\caption{\label{fig:vortexglass} In a vision where positional
order is ignored, and vortices are described by they phases, the
solid phase is transformed into a vortex glass phase, having a
continuous transition (with scaling) towards the liquid phase
\cite{fisher_vortexglass_short,fisher_vortexglass_long}.}
\end{figure}

If one believes that the positional order is lost and dislocations are
spontaneously generated one can wonder whether an elastic
description of the vortex lattice is a good starting point. An
intermediate attitude is to consider that such a description is
useful at intermediate lengthscale and can be used to obtain the
pinning properties (since they are controlled by lengthscales
below $R_a$) or as a first step in absence of a better description
\cite{blatter_vortex_review}. A more radical view is to consider
that since positional order would be lost at large length scale it
is best to ignore it from the start and that an elastic
description of the vortex lattice is a bad starting point: it is
much better to ignore positional order altogether and to focus on
the phase of a vortex
\cite{fisher_vortexglass_short,fisher_vortexglass_long}. The
effect of disorder is thus introduced by a random gauge field
destroying the phase coherence between the vortices. The system is
then described by a random phase energy
\begin{equation}
H = \sum_{ij} \cos(\phi_i - \phi_j - A_{ij})
\end{equation}
With certain assumptions on the properties of the gauge field
(essentially that $\lambda = \infty$) the idea is that the solid
vortex phase will be transformed into a glassy Gauge glass (called
the vortex glass), leading to the phase diagram shown in
Fig.~\ref{fig:vortexglass}.
The vortex glass phase has a continuous transition, with a
divergent lengthscale, towards the liquid. It thus exhibit scaling
at the transition. It was also suggested that inside the glass
phase there should be no linear response to an applied current
\cite{fisher_vortexglass_short}. We will come back to this point
in Section \ref{sec:dynamics}.

Although this description of the vortex lattice/ vortex glass
phase was very successful in the beginning, it started to run into
serious problems as both the experiments and the theory were
refining. Among the experimental problems one would notice (the
corresponding data will be presented in the next sections)
\begin{itemize}
\item The transition between the solid (vortex glass ?) and the
liquid was shown to be discontinuous by various measurements.
Specific heat measurements have now proved that this transition is
first order.

\item Decoration experiments were seeing very large regions free
of dislocations and showing a very good degree of positional
order. This did not seem to fit well with the idea that disorder
would strongly affect the positional order.

\item Neutron scattering was exhibiting quite good Bragg peaks,
again showing stronger positional order than naively anticipated.

\item The phase diagram seemed more complicated than the one shown
in Fig.~\ref{fig:vortexglass}.
\end{itemize}

On the theoretical front two main points were raised: (i) the gauge
glass model was shown to have no glass transition in $d=3$ for
realistic values of the parameters (such as a finite $\lambda$)
\cite{bokil_young_vglass}. (ii) The important suggestion was made,
using scaling arguments \cite{nattermann_pinning,villain_cosine_realrg} and then
firmly established in detailed calculations
\cite{korshunov_variational_short,giamarchi_vortex_short} that
{\it provided dislocations were ignored} displacements in vortex lattices were
growing much more slowly than a power law (logarithmically).

These experimental facts and theoretical points suggested that
the effect of disorder on the vortex lattice could be less destructive than naively anticipated.
They thus strongly prompted for an understanding of the physical properties, such as the positional order,
stemming from the elastic description.
They also made it mandatory to resolve the issue of the asserted \cite{villain_cosine_realrg,fisher_vortexglass_long}
ever presence of disorder induced dislocations, which would invalidate
the elastic results and always destroy the positional order above $R_a$.

\section{Statics : Experimental facts and Bragg glass theory}
\label{sec:statics}

To get a quantitative theory of the disordered system, and go
beyond the simple scaling analysis, it is necessary to
solve the full (\ref{eq:elas}) plus (\ref{eq:dis}). Fortunately
the theoretical ``technology'' had developed tools allowing to
obtain a rather complete solution of this problem
\cite{giamarchi_vortex_short,giamarchi_vortex_long}. We describe
the solution here and examine the consequences for experimental
systems.

\subsection{Bragg glass}

The problem one needs to solve is (using the decomposition of
density (\ref{eq:densexp}))
\begin{equation}
H = \frac{c}2 \int d^dr (\nabla u)^2 + \rho_0 \sum_K \int d^dr
e^{i K (r-u(r))} V(r)
\end{equation}
Although we have written here the simplified form of the elastic
hamiltonian the full one has to be considered but this does not
change the method. One then gets rid of the disorder using
replicas. After averaging over disorder the problem to solve
becomes
\begin{equation} \label{eq:nasty}
H = \frac{c}2 \sum_{a=1}^n \int d^dr (\nabla u_a)^2 - D \rho_0^2
\sum_K \sum_{a,b=1}^n \int d^dr \cos(K (u_a(r)-u_b(r)))
\end{equation}
where $\overline{V(r)V(r')}=D\delta(r-r')$. One has thus traded a
disordered problem for a problem of $n$ interacting fields. The
limit $n\to 0$ has to be taken at the end for the two problems to
be identical. So far the mapping is exact, but (\ref{eq:nasty}) is
still too complicated to be solved exactly. Two methods are
available to tackle it: (i) a variational method; (ii) a
renormalization group method around the upper critical dimension
$d=4$ (a $4-\epsilon$ expansion). The renormalization method is
relatively involved and we refer the reader to the various reviews
and to \cite{giamarchi_vortex_long,emig_exponents_braggglass} for
more details and discussions. The variational method is simple in
principle \cite{feynman_statmech}, and has the advantage to give
the essential physics. One looks for the best quadratic
Hamiltonian
\begin{equation}
H_0 = \sum_{ab} \int d^dq G_{ab}^{-1}(q) u_a(q) u_b(-q)
\end{equation}
that approximate (\ref{eq:nasty}). $H_0$ leads to a variational
free energy
\begin{equation}
F_{\rm var} = F_0 + \langle H - H_0 \rangle_{H_0}
\end{equation}
that has to be minimized with respect to the variational
parameters. The unknown Green's function $G_{ab}(q)$ are thus
determined by
\begin{equation}
\frac{\partial F_{\rm var}}{\partial G_{ab}(q)} = 0
\end{equation}
This is nothing but the well known self consistent harmonic
approximation. The technical complication here consists in taking
the limit $n\to 0$ \cite{mezard_variational_global}. The best
variational parameters are the ones that break the replica
symmetry, in a similar way than what happens in spin glasses. This
is very comforting since we expect on physical grounds that the
competition between the elasticity and the disorder causes a
strong competition where the system has to find its ground state.
It is thus quite natural that such a competition leads to glassy
properties. This is what the solution of the problem confirms. A
similar effect appears in the renormalization solution where a
non-analyticity appears, signaling again glassy properties. The
two methods thus agree quite well (not only qualitatively but also
quantitatively).

\begin{figure}
\centerline{\includegraphics[width=\figwidth]{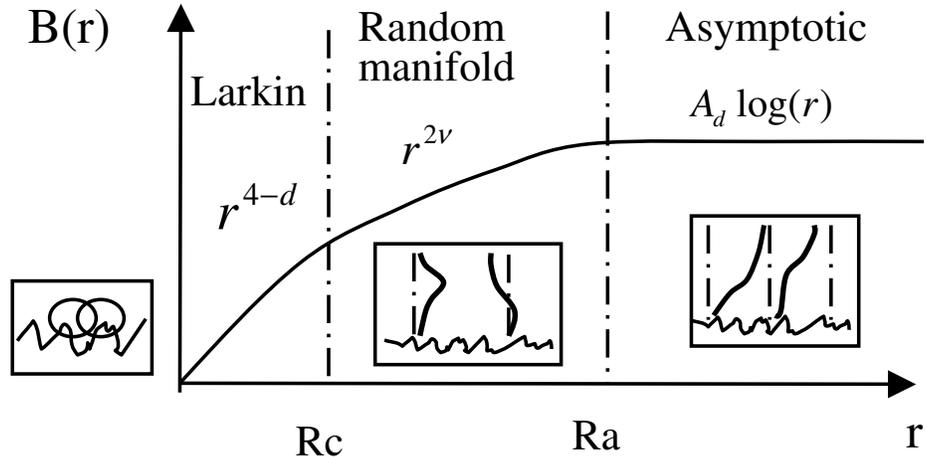}}
\caption{\label{fig:bofr} The relative displacement correlation
function $B(r)$ as a function of distance. The lengthscales $R_c$
and $R_a$ define three regimes. Below $R_c$ one recovers the
Larkin behavior. Between $R_c$ and $R_a$ the displacements still
grow algebraically albeit with a modified exponent. Above $R_a$
the growth becomes logarithmic.}
\end{figure}

\begin{figure}
\centerline{\includegraphics[width=\smallfigwidth]{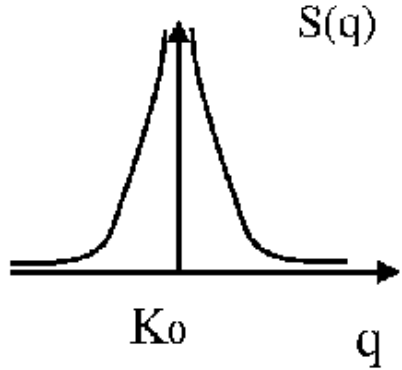}}
\caption{\label{fig:bpeaks} The disordered system still has
quasi-long range positional order and algebraically divergent
Bragg peaks. It is thus nearly as ordered as a perfect solid. Such
a phase, which is a glass when one looks at its dynamics
properties, and in addition possesses perfect topological order
(no defects such as dislocations) is the Bragg glass phase
\cite{giamarchi_vortex_long}.}
\end{figure}

Let us now describe the full solution given in
\cite{giamarchi_vortex_short,giamarchi_vortex_long}. One finds for
the relative displacements correlation function the one shown on
Fig.~\ref{fig:bofr}.
Three regime can be distinguished, separated by the characteristic
lengthscales $R_c$ and $R_a$. Below $R_c$, one is in the Larkin
regime, and the displacements grow as $r^{(4-d)/2}$. Then between
$R_c$ and $R_a$ (i.e. when the displacements are between $\xi$ and
$a$), each line wanders around its equilibrium position and the
problem is very much like the one of a single line in a disordered
environment, i.e. a random manifold problem. The growth of
displacements is still algebraic, albeit with a different exponent
$\nu$. Above $R_a$ however the displacements grow much more slowly
and $B(r) = A_d \log(r)$. The physics is easy to understand:
because of the periodic nature of the system, each line can take
care of the disorder around its equilibrium position. There is
thus no interest for one line to make displacements much larger
than the lattice spacing to pass through a particularly favorable
region of disorder (this would be the case for a single line). For
many lines it is better to let the line closeby take care of this
region. This ensures that the total energy of the {\it whole
system} is the lowest. As a results the displacements do not need
to grow much above $a$. The variational method and RG techniques
allow to compute the prefactor $A_d$ and to address the issue of
the positional order. The positional correlation function is
simply given by
\cite{giamarchi_vortex_short,giamarchi_vortex_long}
\begin{equation}
C(r) \sim  \left(\frac1r\right)^{\eta}
\end{equation}
where $\eta$ is an exponent independent of the strength of
disorder or temperature (for example $\eta_{\rm var} = 1$ in
$d=3$). The physics described above, and obtained from the
variational ansatz is totally generic. The alternate RG approach
in $4-\epsilon$ does indeed recover identical physics, as was
shown first in the Larkin and asymptotic regimes
\cite{giamarchi_vortex_short,giamarchi_vortex_long} and more
recently for the full $B(r)$ \cite{emig_exponents_braggglass}. The
variational approach even gives quite accurate values of the
exponents themselves as can be checked by comparing with the RG.
The quite striking consequence is that far from having the
positional order destroyed by the disorder in an exponential
fashion, a quasi-long range order (algebraic decay of positional
order) exists in the system. Algebraically {\it divergent} Bragg
peaks still exist as shown on Fig.~\ref{fig:bpeaks}.
It is to be noted that this phase, which is thus practically as
ordered as a perfect solid, is a glass when one looks at its
dynamical properties. This is seen in the analysis by the presence
of replica symmetry breaking in the variational approach or the
existence of non-analyticities in the renormalization solution.
From a physical point of view this means that the system has many
metastable states separated from its ground state by divergent
barriers. As a result it exhibits pinning and non-linear dynamics
(creep motion), as we will discuss in more details in
Section~\ref{sec:dynamics}.

What about the argument that dislocations should always be
generated at lengthscale $R_a$ ? In fact it was shown
\cite{giamarchi_vortex_long} that this argument which forgets the
fact that the coupling of the displacements to the disorder is non
linear is simply incorrect and that in fact in $d>2$ dislocations
are {\it not} generated by disorder provided the disorder is
moderate (i.e. $R_a$ large enough compared to $a$). This means
that the results given above corresponds to the {\it true
thermodynamic} ground state of the system. Thus there exists a
thermodynamically stable {\it glassy} phase, the Bragg glass, with
quasi long range order (algebraically divergent Bragg peaks),
perfect topological order (absence of defects such as
dislocations). This is a surprising result since one naively
associate a glass with a very scrambled system. The Bragg glass
shows that this is not the case and that one has to distinguish
the positional properties from the energy landscape (or the
dynamical ones).

The existence of this Bragg glass phase has of course many
consequences for the vortex systems, consequences that we now
investigate and compare with the available experimental data on
vortices.

\subsection{Positional order: Decorations and Neutrons}

The first consequence is the existence of the perfect topological
order and the algebraic Bragg peaks.

\begin{figure}
\centerline{\includegraphics[width=\figwidth]{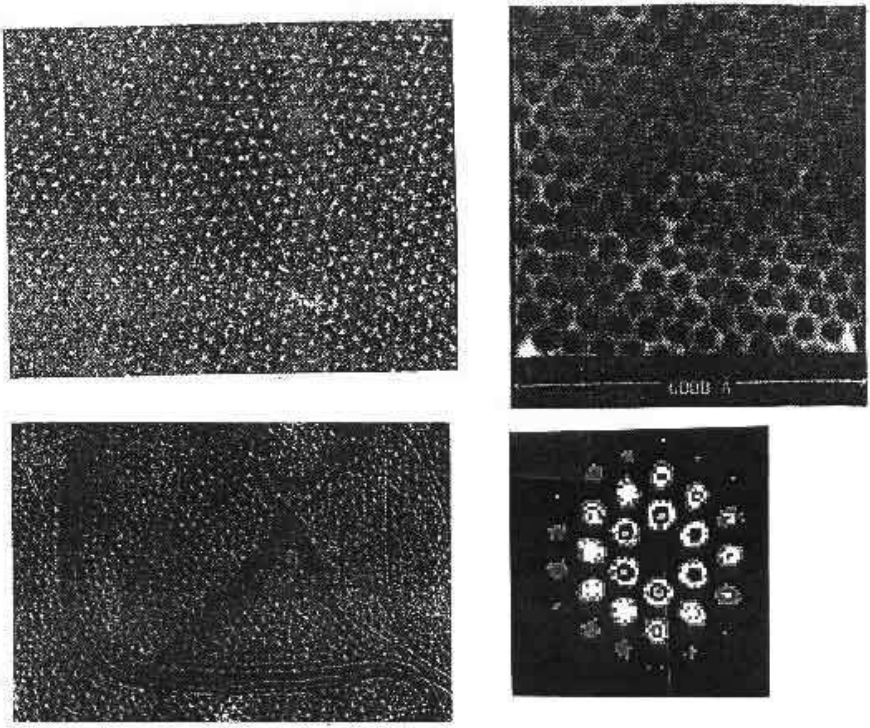}}
\caption{\label{fig:neutnb} Vortex phase structure using various
methods. The upper two panels show a decoration data on the
left\cite{marchevsky_thesis} and an STM data on the right in
NbSe2\cite{hess_stm_vortex}. The lower two panels show a Lorentz
holography micrograph of a thin film
Nb\cite{harada_lorentz_vortex} and a neutron diffraction picture
of a single crystal Nb\cite{ling_neutrons_bragg} }
\end{figure}

On the theoretical side, after the initial proposal
\cite{giamarchi_vortex_long} the existence of algegraic bragg
peaks and the absence of dislocations have been confirmed by
further analytical results
\cite{carpentier_bglass_layered,kierfeld_bglass_layered,fisher_bragg_proof}
and numerical simulations
\cite{gingras_dislocations_numerics,vanotterlo_bragg_numerics}. On
the experimental side, direct structural information has been
obtained from both real space studies such as magnetic decoration,
scanning tunneling microscopy, Lorentz microscopy and reciprocal
space studies by neutron diffraction, summarized in
Fig.~\ref{fig:neutnb}. The upper two panels show a decoration
micrograph of a fairly ordered lattice in NbSe2 with a few defects
\cite{marchevsky_thesis} and an STM micrograph
\cite{hess_stm_vortex} of the same material with no defects. The
Lorentz micrograph in a thin film of Nb
\cite{harada_lorentz_vortex} produces a nearly amorphous vortex
assembly while the neutron diffraction data on a single crystal
sample of Nb \cite{ling_neutrons_bragg} show Bragg peaks up to
third order reflections, suggesting a very high degree of order.
These results show that defect-free phases with Bragg reflections
are experimentally observed, while highly defective or even
amorphous or liquid like phases also exist. The task is to find
which parts of the phase space are occupied by each and what
controls the phase transformations among them.

The fourier transforms of the real space data
\cite{grier_decoration_manips,marchevsky_thesis,kim_decorations_nbse}
show very large regions free of dislocations and yield Bragg
peaks routinely, suggesting a much stronger solid like order than
would be expected naively from a vortex glass model. The situation
with neutron diffraction is similar as shown on
Figure~\ref{fig:neutnb}.

\begin{figure}
\centerline{\includegraphics[width=\smallfigwidth]{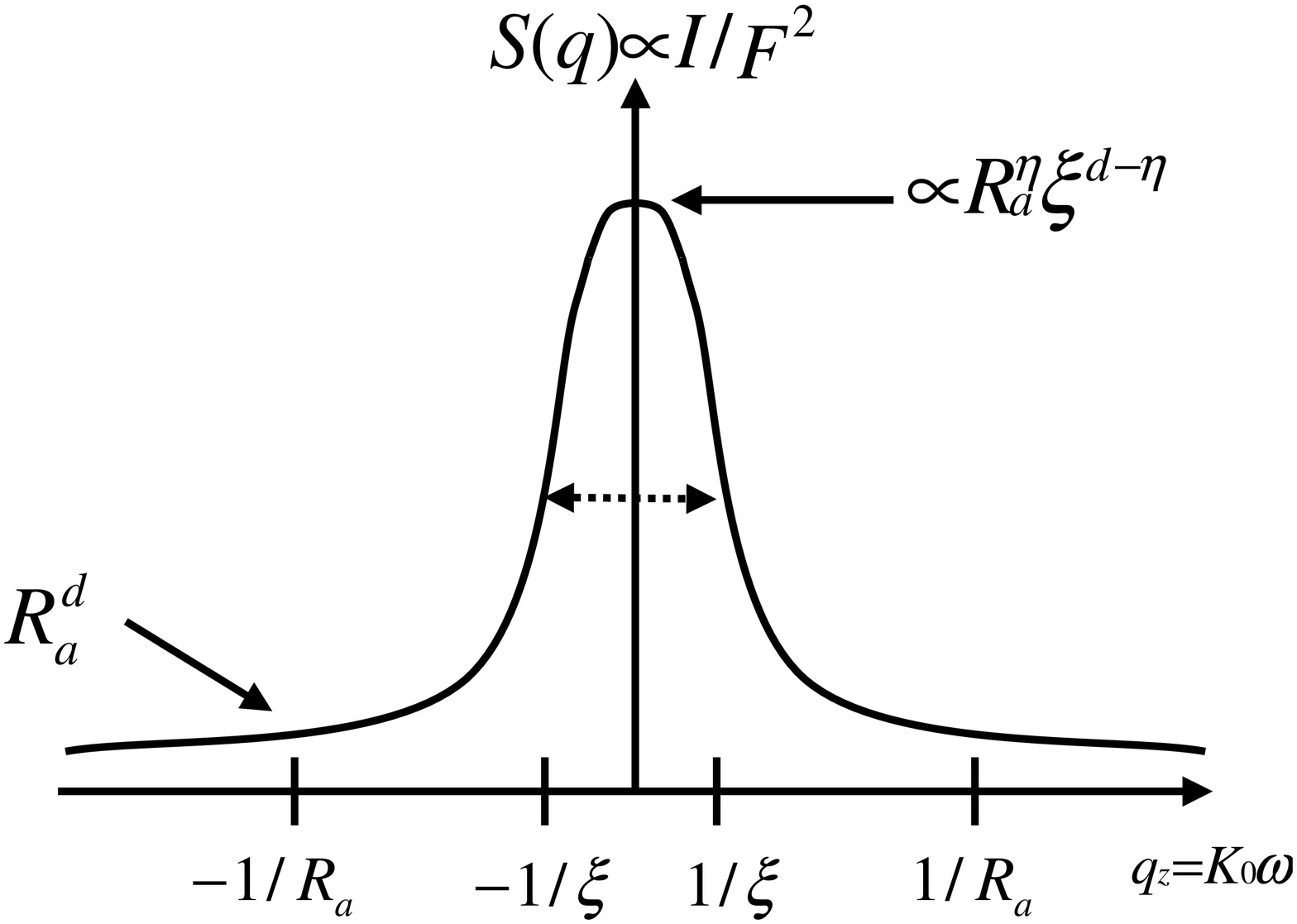}
            \includegraphics[width=\smallfigwidth]{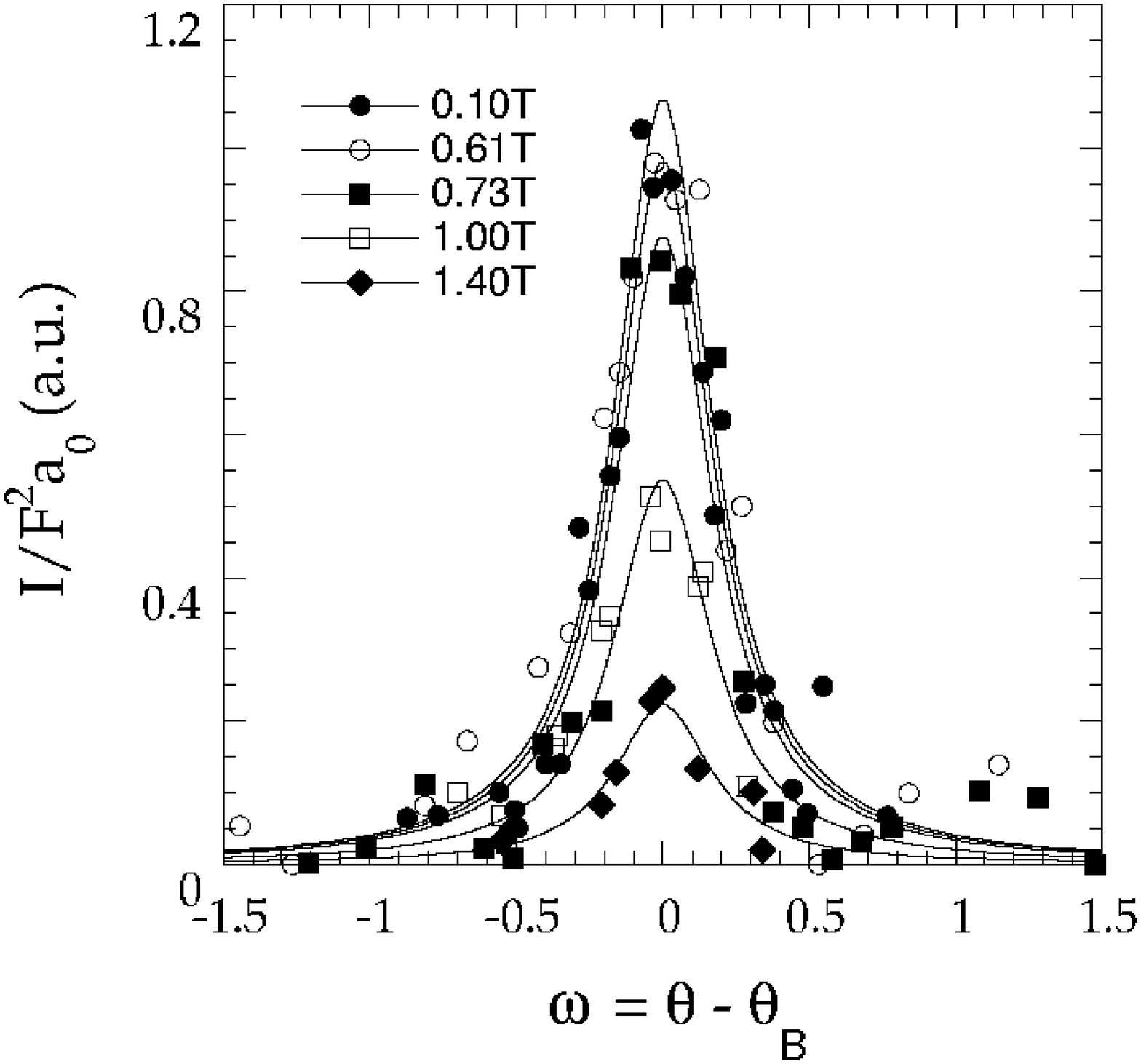}}
\caption{\label{fig:neutronbkbo} (left) Bragg glass predictions
for the angular dependence of the neutron diffracted intensity
when a finite experimental resolution is taken into account. The
arrows indicate the values of $S(q)$ for $q=0$ and $1/R_a$
respectively. If the disorder increases (Ra decreases), the height
of the peak decreases as $R_a^\eta$ but the peak does not broaden
since the half width at half maximum is always given by the
experimental resolution $1/\xi$ and not $1/R_a$. The height of the
peaks gives a direct measure of the characteristic length
$R_a$.(right) Angular dependence of neutron intensity diffracted
by a (K,Ba)BiO$_3$ crystal \cite{klein_brglass_nature} at the
indicated applied fields. Those data show that the diffracted
intensity (rocking curves) collapse without any broadening above
$~0.7T$.}
\end{figure}

Recently neutron data have provided a direct evidence of the presence of the Bragg glass phase.
Indeed the power-law Bragg peaks shown in Fig.~\ref{fig:bpeaks} gives when convolved with
the experimental resolution the result shown in Fig.~\ref{fig:neutronbkbo}. The width at half width of the
observed peak is constant and determined by the experimental resolution, and the height is fixed by the position
correlation length $R_a$. Thus if disorder (or magnetic field -- see in the next section) is increased the
Bragg glass predicts that observed neutron peaks should collapse without broadening.
This behavior, has been quantitatively tested on the compound (K,Ba)BiO$_3$ \cite{klein_brglass_nature}
(see Fig.~\ref{fig:neutronbkbo}) which has the advantage of being totally isotropic and thus avoid
all complications associated with anisotropy such as possible 2D-3D crossovers. Peaks are seen to collapse
without any broadening thus providing a direct evidence of the Bragg glass phase and its algebraic positional
order.

\subsection{Unified phase diagram}

\begin{figure}
\centerline{\includegraphics[width=\medfigwidth]{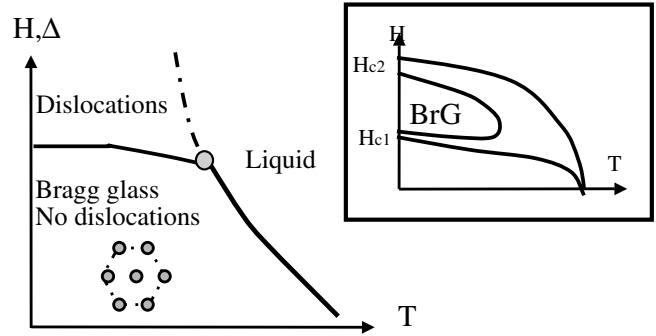}}
\caption{\label{fig:uniphas} Universal phase diagram for type II
superconductors. The Bragg glass ``melts'' due to thermal
fluctuations of disorder induced fluctuations. This lead to
transitions as a function of magnetic field. The left diagram is
far from $H_{c1}$ (more adapted to High Tc) whereas the insert
shown the full diagram. The melting towards the liquid is expected
to be first order.}
\end{figure}

Another striking consequences of the existence of the Bragg glass,
is that it imposes a {\it generic} phase diagram for all type II
superconductors \cite{giamarchi_vortex_long}. Indeed the Bragg
glass has no dislocations, thus if either thermal fluctuations
{\it or} disorder are increased the Bragg glass should ``melt'' to
a phase that contains defects . If thermal fluctuations increase
this is the standard melting and leads to the liquid phase.
Because the Bragg glass is nearly as ordered as a perfect solid
one can expect it to melt though a first order phase transition.
More importantly this ``melting'' of the Bragg glass can also
occur because the disorder is increased in the system. For
vortices increasing the field has a similar effect. Indeed the
effective disorder in (\ref{eq:dis}) is $V \rho_0$, thus
increasing the average density makes the disorder term stronger
compared to the elastic term (\ref{eq:elas}). Indeed for moderate
fields the change in elastic constants due to the field is quite
small. Thus increasing the field is like increasing disorder. One
should thus have a {\it ``melting''} transition of the Bragg glass
(induced by the disorder) as a function of the field. Close to
$H_{c1}$ because the change of elastic constants is then drastic,
one expects a similar transition. This leads to the universal
phase diagram shown in Fig.~\ref{fig:uniphas}.

\begin{figure}
\centerline{\includegraphics[width=\figwidth]{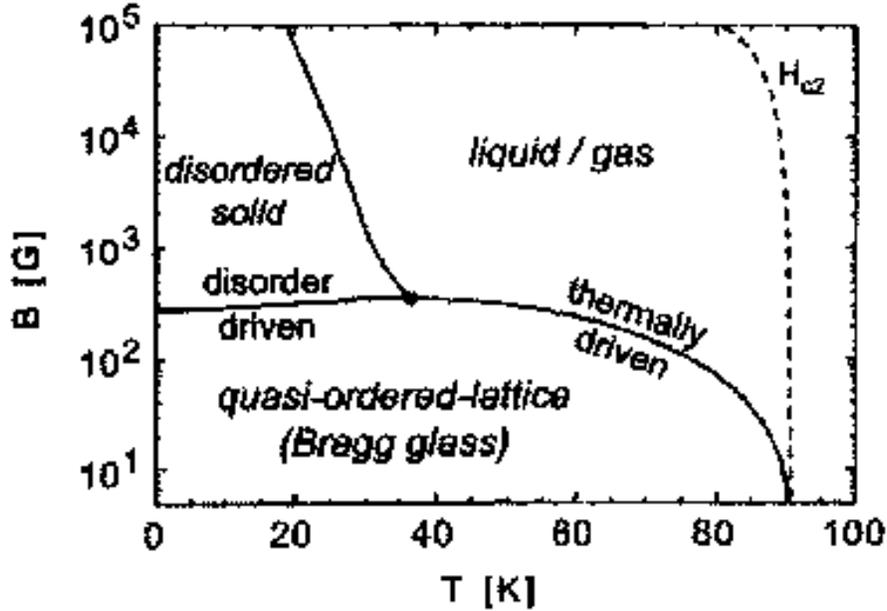}}
\caption{\label{fig:expdiag} Experimental phase diagram for the
cuprate system BSCCO
\cite{zeldov_diagphas_bisco,khaykovich_diagphas_bisco}. At the
thermally induced melting transition one sees thermodynamic
anomalies as in Fig.~\ref{fig:jumps}. The disorder-driven
transition is accompanied by the second peak or peak effect, shown
later in Fig.~\ref{fig:peakstructure}.}
\end{figure}

\begin{figure}
\centerline{\includegraphics[width=\medfigwidth]{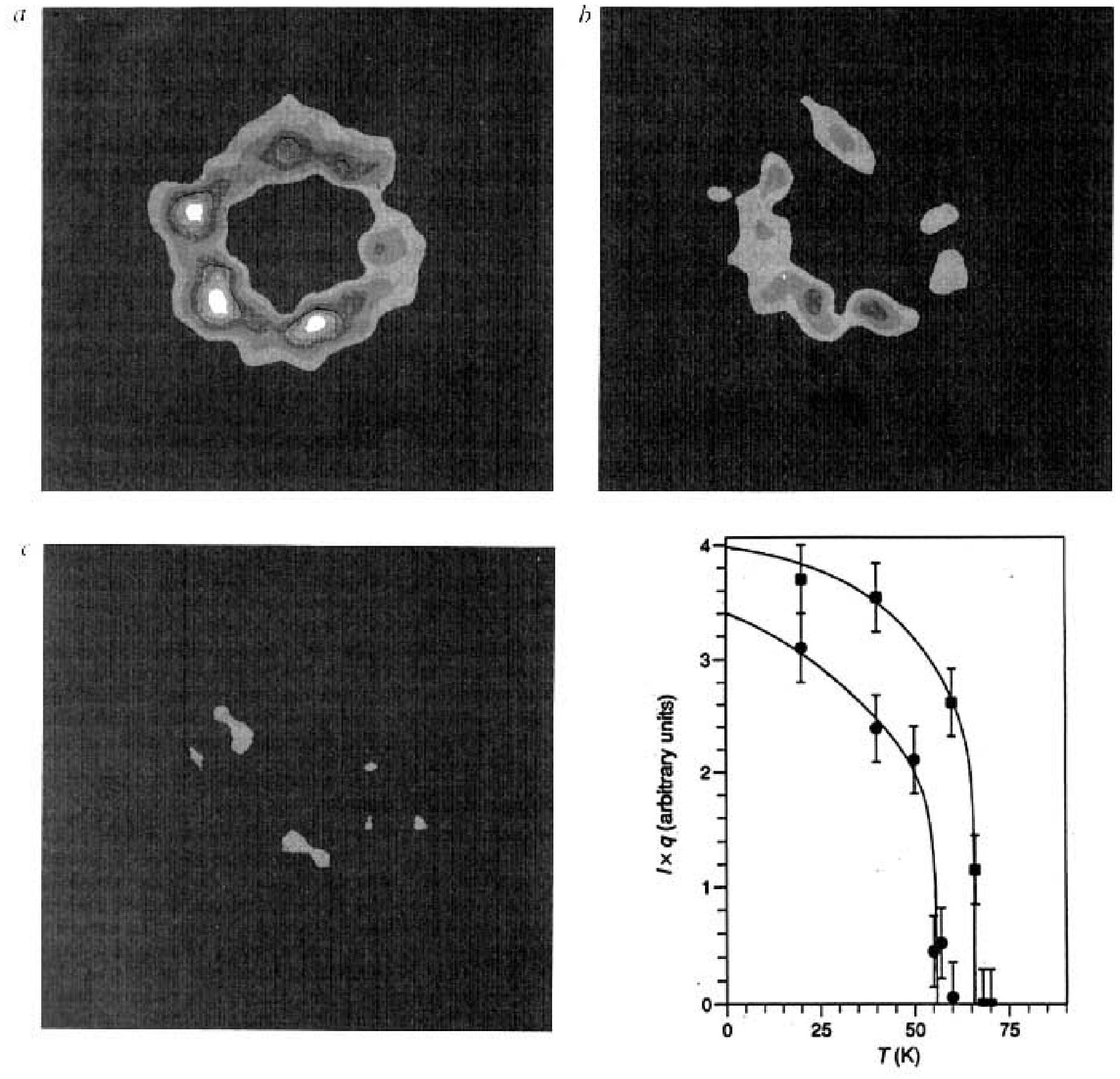}}
\caption{\label{fig:neutron1} Evolution of the neutron bragg peaks
in the ordered phase (upper left panel) with increasing
temperature. The lower right panel shows the loss of Bragg
intensity with increasing T}
\end{figure}

\begin{figure}
\centerline{\includegraphics[width=\medfigwidth]{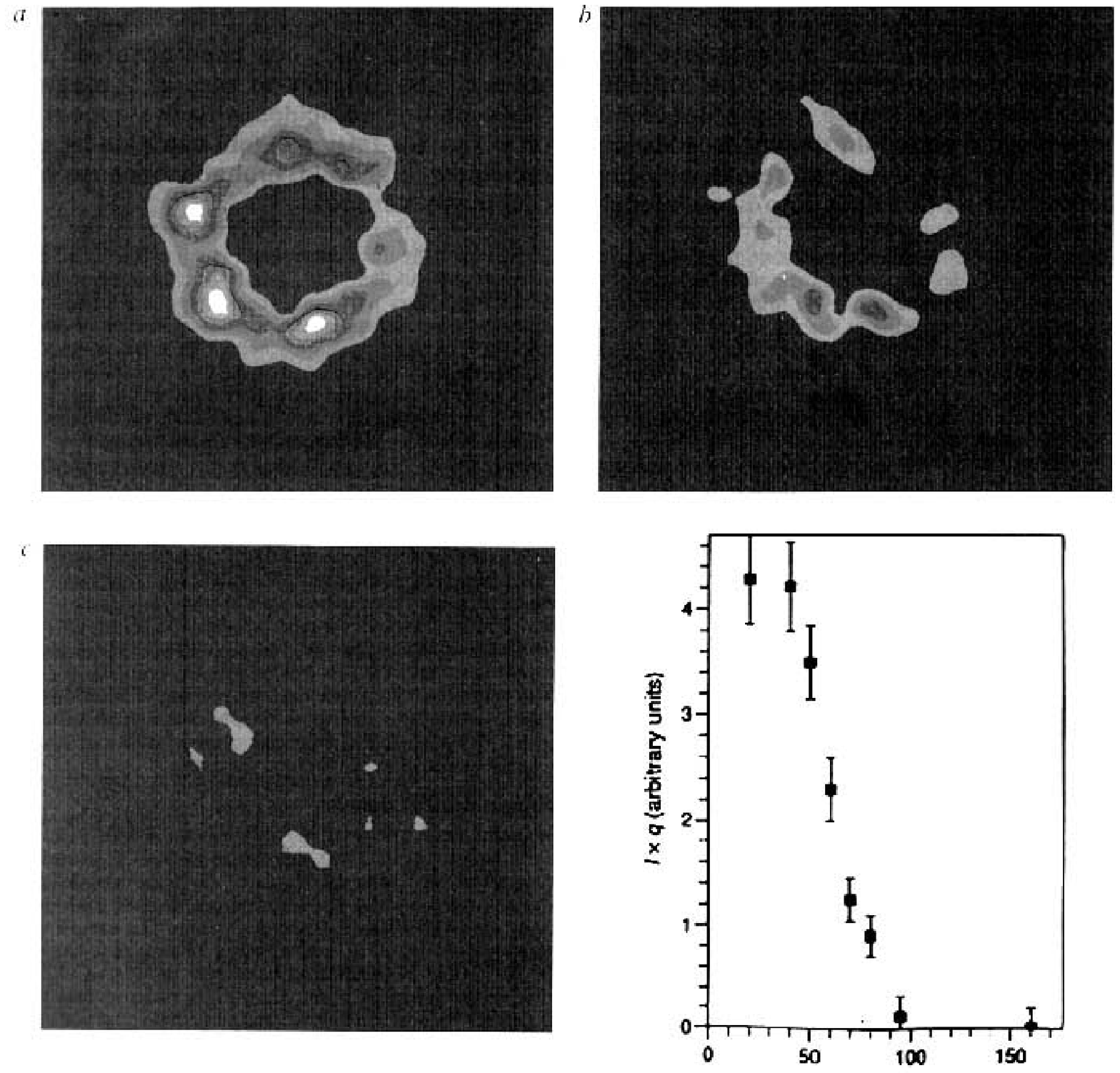}}
\caption{\label{fig:neutron2} Loss of Bragg intensity upon
increasing the magnetic field at a fixed temperature}
\end{figure}

\begin{figure}
\centerline{\includegraphics[width=\figwidth]{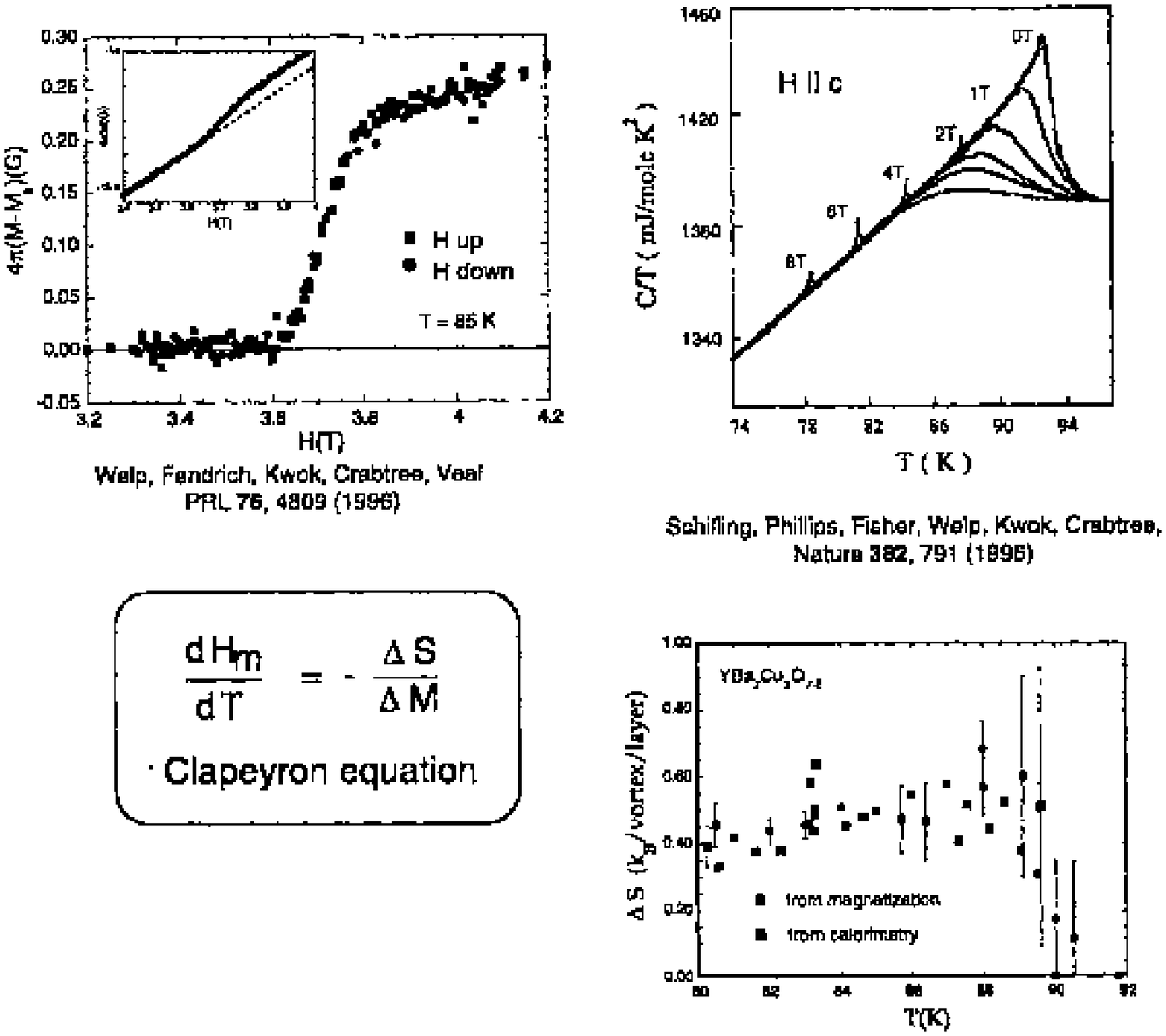}}
\caption{\label{fig:composite} Experimental data
\cite{welp_magnetization_jump,schilling_heat_vortex} showing
thermodynamic measurements from magnetization and calorimetry of
the first order transition in YBCO}
\end{figure}

We first focus on the experimental results of the loss of Bragg
glass order in the cuprate systems. A remarkable experimental
determination of the phase diagram of BSCCO is shown in
Fig.~\ref{fig:expdiag}.
Three phases are clearly identified in it : a quasi-ordered Bragg
glass phase which ``melts'' by thermal fluctuations into a liquid
and also ``amorphizes'' or ``melts'' by quenched disorder into a
disordered solid. First we focus on structural evidence of the two
phase transitions from neutron diffraction
\cite{cubbit_neutrons_bscco}. In the Bragg glass phase one clearly
observes the resolution limited Bragg peaks. Upon increasing
temperature across the thermally driven transition the Bragg peaks
lose intensity and become unobservable in the putative liquid
phase. A similar loss of Bragg peak intensity is observed when the
magnetic field is increased at fixed $T$ across the Bragg glass to
the disordered or vortex glass phase. Fig.~\ref{fig:neutron1} and
Fig.~\ref{fig:neutron2} show the results of neutron diffraction of
BSCCO as the two phase boundaries are crossed, one from the
ordered Bragg glass to the liquid and the other from Bragg glass
to the disordered solid phase \cite{cubbit_neutrons_bscco}.
In both cases the Bragg reflections lose intensity and disappear
at the phase boundary, in a way similar to the one discussed in
Fig.~\ref{fig:neutronbkbo}. Due to limited neutron intensity,
detailed studies of the disordered phase has not been performed in
the cuprates. But recent neutron diffraction studies of low Tc
system Nb \cite{gammel_neutrons_Nb,ling_neutrons_bragg} with a
short penetration depth have directly shown a transformation of
bragg peaks to a ring of scattering, implying liquid like
(amorphous) order in the disordered phase. The experimental phase
behavior is thus entirely compatible with theoretical expectations
in Fig.~\ref{fig:uniphas}. The comparison with the theoretical
phase diagram identifies the quasi-ordered solid phase with the
Bragg glass phase. The phase with dislocations is expected to
correspond to the disordered solid phase. For BSCCO the position
of the field melting line has been computed by Lindemann argument
or similar cage arguments
\cite{ertas_diagphas_bisco,giamarchi_diagphas_prb,kierfeld_diagphas_bisco,koshelev_diagphas_bisco}
and the value of the ``melting'' field is in good agreement with
the observed experimental value. The distinction between the
disordered solid phase and the liquid phase remains an
experimentally open question for different systems with different
types of disorder. For the very anisotropic BSCCO system there is
also the question of the existence of additional phases.
Structural results of the same kind are not available for the
other common cuprate system, namely YBCO. However, thermodynamic
data on the magnetization jump and entropy jump were measured. A
composite of the data is shown in Fig.~\ref{fig:composite} (see also
\cite{bouquet_melting_ybco}) which
demonstrates close agreement within the Clapeyron equation
confirming the first order nature of the thermally driven melting
transition.

\subsection{Second peak and peak effect} \label{sec:peakeffect}

\begin{figure}
\centerline{\includegraphics[width=\medfigwidth]{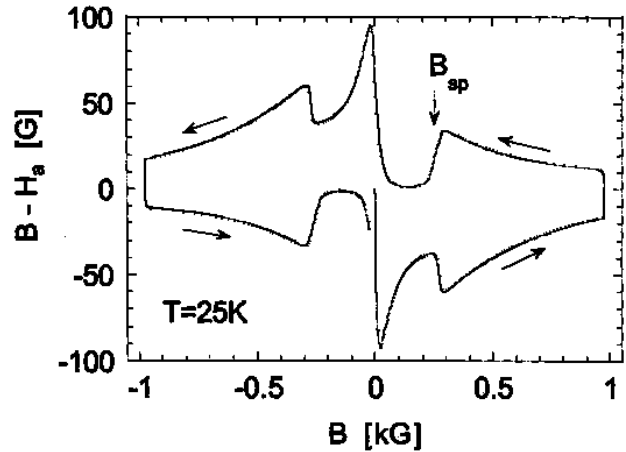}}
\caption{\label{fig:fishtail} Typical second magnetization peak
data from the magnetic hysteresis loop in
BSCCO\cite{khaykovich_diagphas_bisco}, the sudden enhancement of
the diamagnetic signal at $B_{sp}$ marks the enhancement of
critical current across the Bragg glass to vortex glass phase
transition.}
\end{figure}

Now we focus on the disorder driven phase transition shown above.
The magnetization hysteresis loop across this transition manifests
a typical second peak effect, often called the fishtail effect due
to its shape, as shown in Fig.~\ref{fig:fishtail} for BSCCO
\cite{khaykovich_diagphas_bisco}.
The sharp jump is magnetization at $B_{sp}$ marks a sudden
increase in the irreversibility, i.e., a jump in critical current
$J_c$. In a naive view of the Larkin scenario, this marks a sudden
decrease in the correlation volume, i.e., a sharp loss of order,
consistent with the neutron diffraction data shown above. Similar
results were obtained for YBCO also, yielding a qualitatively
similar phase diagram
\cite{deligiannis_diagphas_ybco,nishizaki_diagphas_ybco,kokkaliaris_diagphas_ybco}.

\begin{figure}
\centerline{\includegraphics[width=\figwidth]{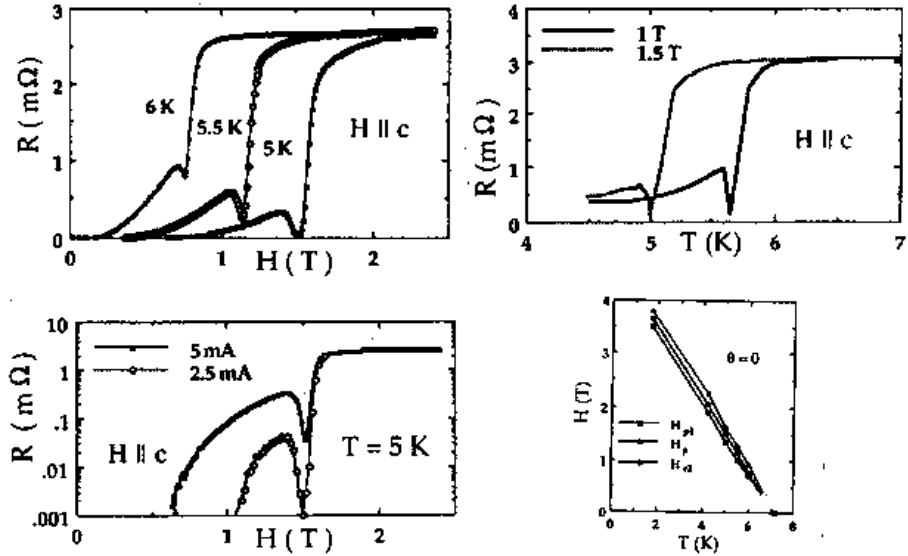}}
\caption{\label{fig:peakeffect} Typical observation of peak effect
in transport \cite{higgins_second_peak}. The upper two panels show
the transition in isothermal and isofield measurements. The lower
left panel shows the independence of the locus on driving current
suggesting a thermodynamic origin of the anomaly. The right panel
shows the locus of the peak region bounded by the onset at
$H_{pl}$ and the peak at $H_p$. The close proximity to the upper
critical field is typical of low $T_c$ materials..}
\end{figure}

\begin{figure}
\centerline{\includegraphics[width=\figwidth]{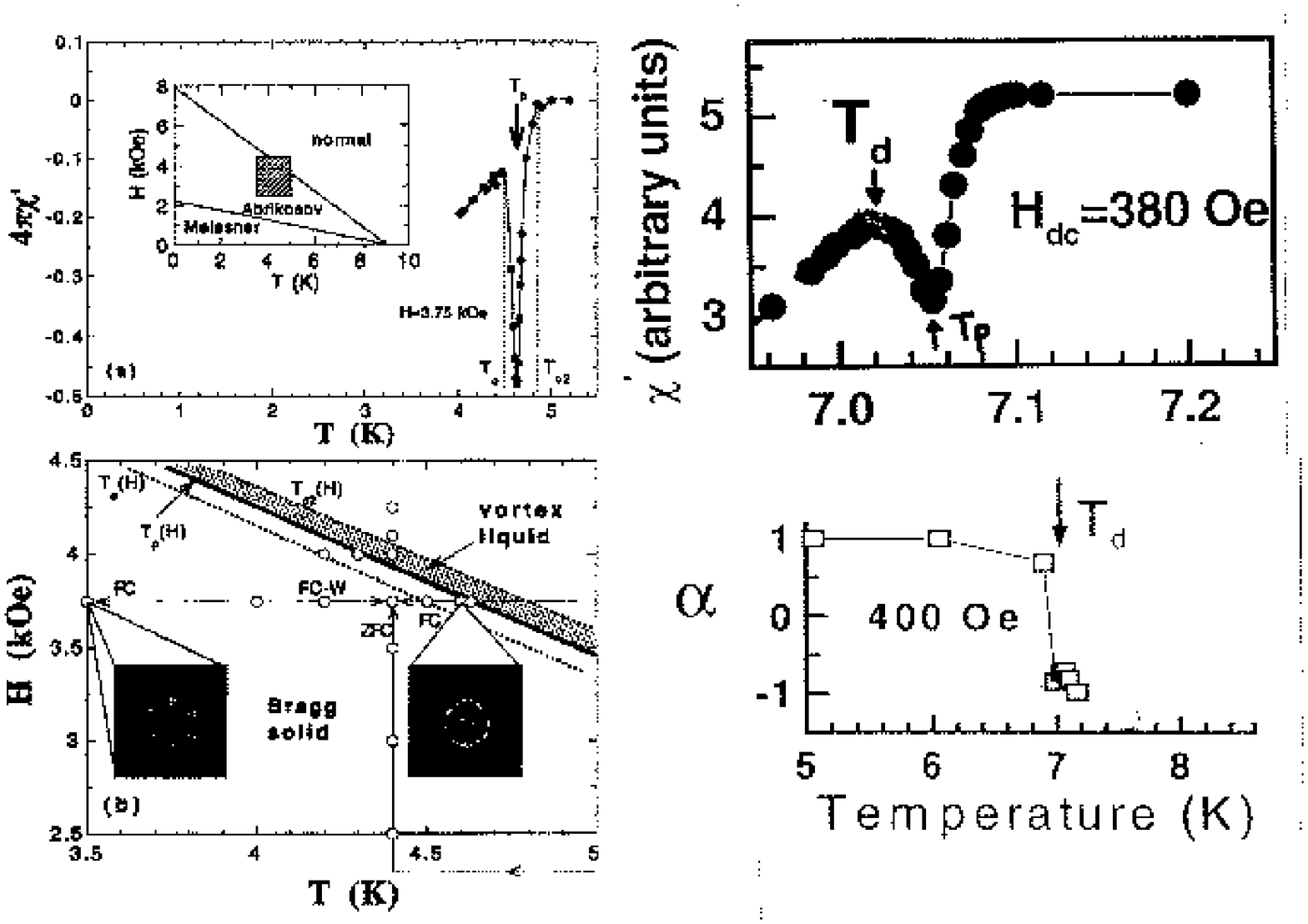}}
\caption{\label{fig:peakstructure} Experimental data on
Nb\cite{ling_neutrons_bragg} on the left and NbSe2
\cite{rao_musr_peak} on the right. The upper panel shows the peak
effect from ac susceptibity for both. The lower left panel shows
the structure from neutron diffraction in Nb showing the Bragg
spots disappear into a ring of scattering. Muon spin relaxation
data in NbSe2 on the right shows a sudden drop in the skewness
parameter at the peak effect that marks the first order structural
disordering of the lattice.}
\end{figure}

Peak effects are ubiquitous in low Tc systems and have been known
for nearly four decades and provided the primary motivation for
the Larkin-Ovchinnikov scenario of collective pinning. In these
systems the peak effect \cite{wordenweber_kes_peak}  usually occurs very close to the normal
phase boundary, unlike in the cuprate systems where the fishtail anomaly occurs very far from it.
Recent resurgence
of activity on the peak effect phenomenon in low Tc systems also
provide a phase diagram of ordered and disordered vortex phases not dissimilar to the cuprates.
Due to the smallness of the
thermal fluctuation effects, however, the peak effect transition
often occurs in close proximity to the melting transition, or even coincident with it.
Separating the two effects remains a matter of
considerable controversy in the low Tc systems. In the popular low
Tc system NbSe2 the peak effect phenomenon has been studied
extensively in recent years \cite{bhattacharya_peak_prl}. Fig.~\ref{fig:peakeffect} shows a typical set of data for the
resistive detection of the peak effect where a rapid drop in the
resistance at the peak effect boundary marks a sudden increase in
the critical current analogous to the magnetization jump shown
above.
Direct structural evidence also clearly shows the amorphization of
the Bragg glass phase with six fold symmetric Bragg spots to a
ring of scattering at the peak in the elemental system Nb
\cite{ling_neutrons_bragg}. In NbSe2 the same peak effect is
accompanied by a sharp change in the line shape as seen in the
asymmetry parameter \cite{rao_musr_peak}. These results are shown
in Fig.~\ref{fig:peakstructure}.

\begin{figure}
\centerline{\includegraphics[width=\figwidth]{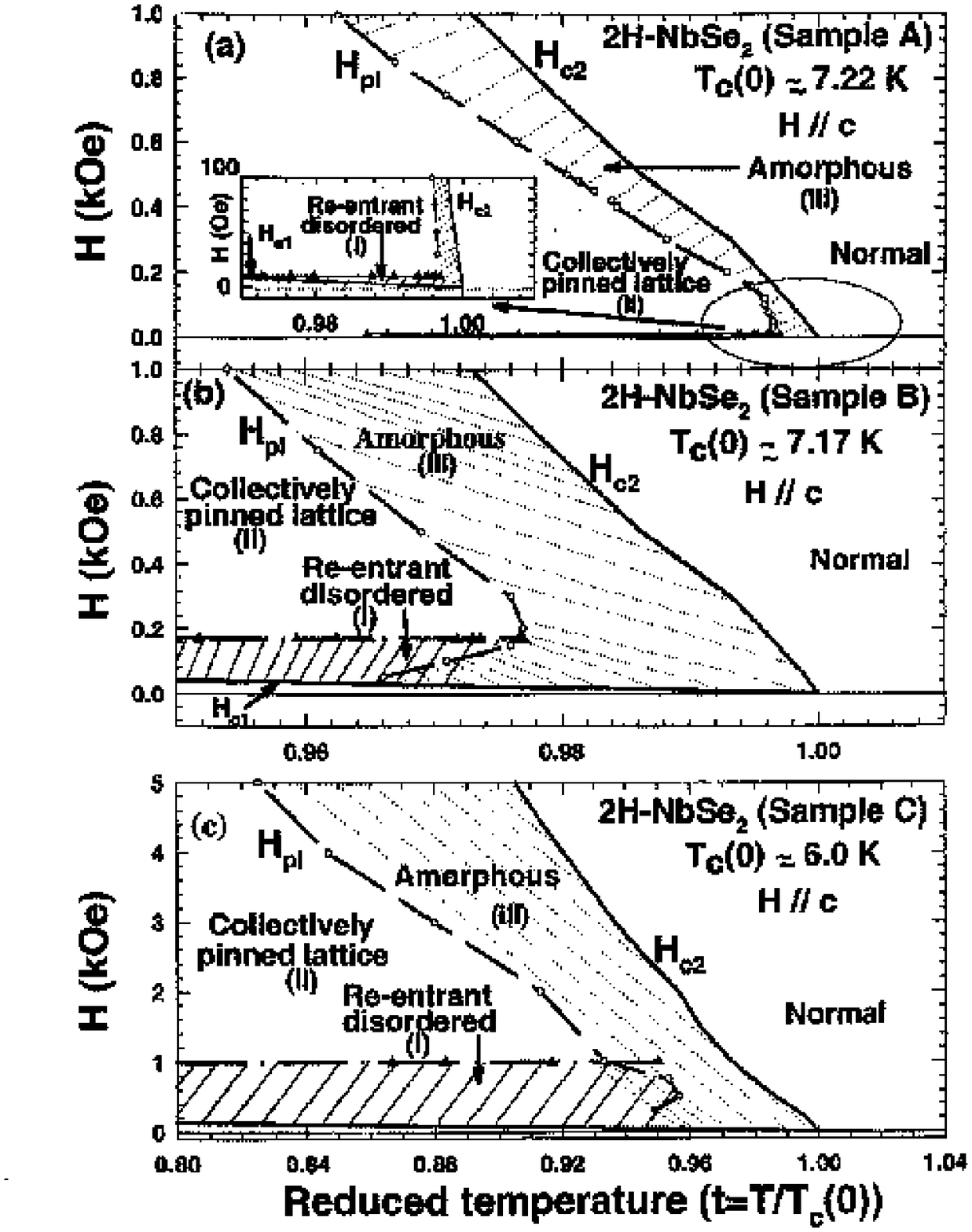}}
\caption{\label{fig:shrink} Evolution of the phase diagram
\cite{banerjee_reentrant_peak} with increased disorder from sample
A to C. As disorder increases, the collectively pinned lattice,
i.e., the Bragg glass phase, shrinks from both above and below.}
\end{figure}

Of special importance is the clear
experimental observation of a reentrant phase behavior for the
NbSe2 system \cite{ghosh_reentrant_peak,banerjee_reentrant_peak}.
From the Meissner phase an increase in field shows
two anomalies, first from a disordered phase into an ordered phase
and then a reentry into a disordered phase just below the upper
critical field. Especially significant is the pronounced shift in
the order-disorder phase boundary with varying quenched disorder
(addition of magnetic dopants). Figure~\ref{fig:shrink} shows the
progressive shrinkage of the Bragg glass phase from both high
field side as well as from the low field side as disorder is
increased from sample A through sample C.
These results are in
excellent agreement with the theoretical discussions above. Direct
structural evidence of amorphization in this system was obtained
through muon spin relaxation experiments \cite{rao_musr_peak} that are entirely
analogous to the results \cite{aegerter_YBCO_musr} in the cuprate systems.

Several questions remain open for the peak effect from the phase
behavior point of view. In addition to the second magnetization
peaks, a peak effect is often observed in YBCO very close to or
even coincident with the melting transition
\cite{ishida_peak_melting}. This suggests that the disordered
solid phase may protrude as a sliver all around the Bragg glass
phase \cite{menon_phasediag_univ} or that there are two types of
peak effects, one associated with disorder induced melting and the
other with the thermally induced melting transition. In what
follows, we indeed show that the melting of the Bragg glass
provides a natural explanation for the peak effect.

\begin{figure}
\centerline{\includegraphics[width=\smallfigwidth]{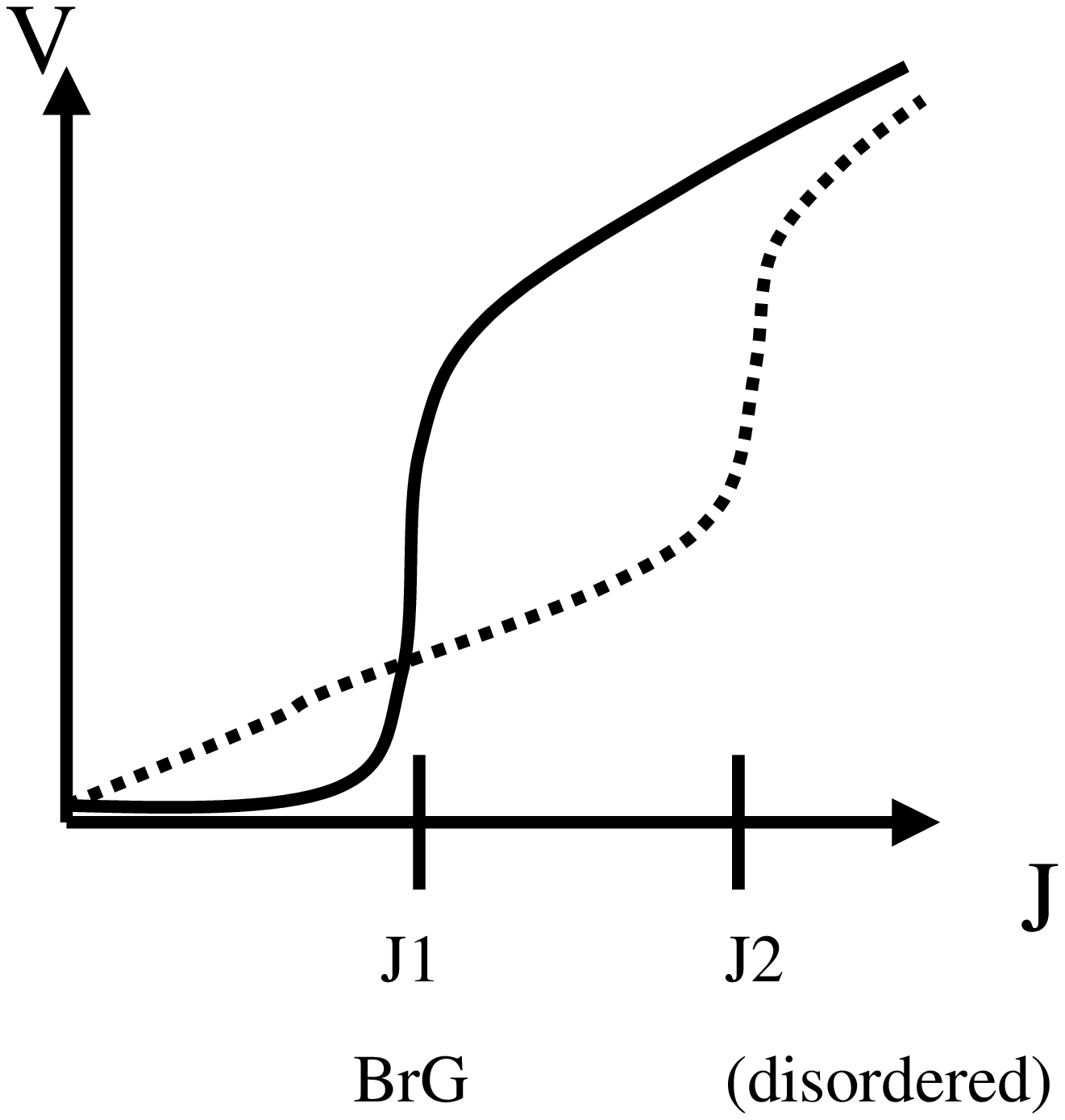}}
\caption{\label{fig:peak} The $V-I$ characteristics for the Bragg
glass (full line) and the high field phase or the liquid (dashed
line) \cite{giamarchi_diagphas_prb}.}
\end{figure}

How the melting of the Bragg glass signals itself ? To understand
it let us look at the $V-I$ characteristics, shown on
Fig.~\ref{fig:peak}.
The Bragg glass is collectively pinned so it has a small critical
current $J_1$ but very high barriers leading to practically no
motion (hence no $V$) in the pinned phase (below $J_1$). On the
other hand in the high field phase (with dislocations) or the
liquid it is more easy to pin small parts leading to higher
critical currents, but the pinning is not collective hence a much
more linear response below $J_2$. The $V-I$ characteristics thus
cross at the melting of the Bragg glass
\cite{giamarchi_diagphas_prb}. This crossing leads for an apparent
increase of the critical current close to the melting and thus to
a peak effect in the transport measurements or to a second peak in
the magnetization measurements.

Recent simulations and reexamination of older experimental data
\cite{vanotterloo_IV_peakeffect,higgins_second_peak} are in excellent
agreement with this $I-V$ crossing scenario at the peak effect near
melting or the second magnetization peak.

\section{Dynamics of vortices} \label{sec:dynamics}

\begin{figure}
\centerline{\includegraphics[width=\figwidth]{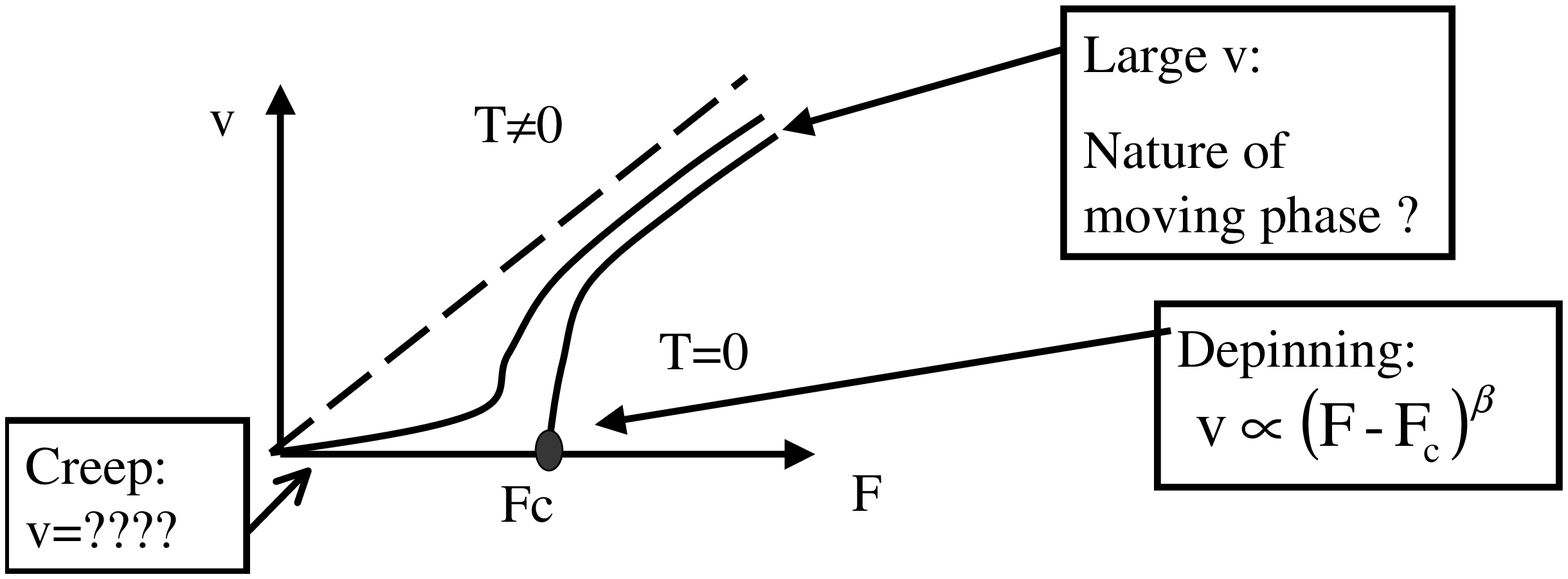}}
\caption{\label{fig:vi} The $V-I$ characteristics for a vortex
system. This is in fact the velocity of the system $v$, in
response to an external force $F$.}
\end{figure}

The competition between disorder and elasticity manifests also in
the dynamics of such systems, and if any in a more dramatic
manner. When looking at the dynamics, many questions arise. Some
of them can be easily asked (but not easily answered) when looking
at the $V-I$ characteristics shown in Fig.~\ref{fig:vi}.
The simplest question is prompted by the $T=0$ behavior. Since the
system is pinned the velocity is zero below a certain critical
force $F_c$. For $F>F_c$ the system moves. What is $F_c$ ? We saw
that the scaling theory of Larkin and Ovchinikov relates it
directly to the static characteristic length $R_c$. Can one
extract this critical force directly from the solution of the
equation of motion of the vortex lines
\begin{equation}
\eta \frac{d u_i(t)}{dt} = - \frac{\delta H}{\delta u_i} + F +
\zeta_i(t)
\end{equation}
This equation is written for overdamped dynamics, but can include inertia as well.
$\eta$ is the friction coefficient taking into account the dissipation in the cores,
$F$ the externally applied force, and $\zeta$ a thermal noise.
Can one obtain from this equation the velocity above $F_c$ ? The $v-F$ curve at
$T=0$ is reminiscent of the one of an order parameter in a second
order phase transition. Here the system is out of equilibrium so
no direct analogy is possible but this suggests that one could
expect $v \sim (F-F_c)^\beta$ with a dynamical critical exponent
$\beta$. We will not investigate these questions here, because of
lack of space and refer the reader to the above mentioned reviews and
to \cite{chauve_creep_long} for an up to date discussion of this issue
and additional references.

The second question comes from the $T\ne 0$ curve. Well below
threshold $F \ll F_c$ the system is still expected to move through
thermal activation. What is the nature of this motion and what is
the velocity ? If the system has a glassy nature one expects it to
manifest strongly in this regime since thermal activation will
have to overcome barriers between various states. We address this
question in Section~\ref{sec:creep}.

Finally, there are many questions beyond the simple knowledge of
the average velocity. One of the most interesting is the nature of
the moving phase. If one is in the moving frame where the system
looks motionless, how much this moving system resembles or not the
static one. This concerns both the positional order properties and
the fluctuations in velocity such as the ones measured in noise
experiments. This is discussed in Section~\ref{sec:dynamical}.

\subsection{Creep} \label{sec:creep}

\begin{figure}
\centerline{\includegraphics[width=\smallfigwidth]{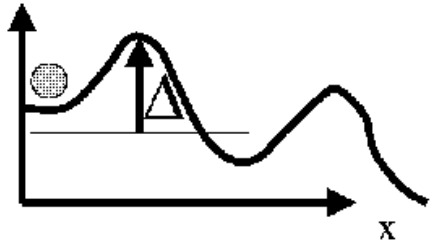}}
\caption{\label{fig:landscape} In the Thermally Assisted Flux Flow
(TAFF) \cite{anderson_kim} a region of pinned material is
considered as a particle moving in an energy landscape with
barriers. This leads to an exponentionally small but linear
response.}
\end{figure}

Let us first examine the response of the vortex system to a very
small external force. For usual systems we expect the response to
be linear (leading to a finite resistivity for the system). Indeed
earlier theories of such a motion found a linear response. The
idea is to consider that a blob of pinned material has to move in
an energy landscape with barriers $\Delta$ as shown in
Fig.~\ref{fig:landscape}.
The external force $F$ tilts the energy landscape making forward
motion possible. The barriers are overcome by thermal activation
(hence the name: Thermally Assisted Flux Flow) with an Arrhenius
law. If the minima are separated by a distance $a$ the velocity is
\begin{equation}
v \propto e^{-\beta(\Delta - Fa/2)} - e^{-\beta(\Delta + Fa/2)}
\simeq e^{- \beta \Delta} F
\end{equation}
The response is thus linear, but exponentially small. One thus
recovers that pinning drastically improves the transport qualities
of superconductors. For old superconductors $\beta$ was small
enough so that this formula was not seriously challenged. However
with high $T_c$ one could reach values of $\beta$ where it was clear
that the TAFF formula was grossly overestimating the motion of
vortex lines.

The reasons is easy to understand if one remembers that the static
system is in a vitreous state. In such states a ``typical''
barrier $\Delta$ does not exist, but barriers are expected to
diverge as one gets closer to the ground state of the system. The
TAFF formula is thus valid in system where the glassy aspect is
killed. This is the case in the liquid where various parts of the
system are pinned individually. When the glassy nature of the
system persists up to arbitrarily large lengthscales the theory
should be accommodated to take into account the divergent
barriers \cite{fisher_vortexglass_short,fisher_vortexglass_long}.
This can be done quantitatively within the framework of the elastic
description. In fact such a theory was developed before for interfaces
\cite{nattermann_rfield_rbond,ioffe_creep} and then
adapted for periodic systems such as the vortex lattice \cite{feigelman_collective,nattermann_pinning}.
The basic idea is beautifully simple. It rests on two quite
strong but reasonable assumptions : (i) the motion is so slow that
one can consider at each stage the lattice as motionless and use
the {\it static} description; (ii) the scaling for barriers which
is quite difficult to determine is the same than the scaling of
the minimum of energy (metastable states) that can be extracted
again from the static calculation. If the displacements scale as
$u \sim L^\nu$ then the energy of the metastable states (see
(\ref{eq:elas})) scale as
\begin{equation} \label{eq:scalmet}
E \sim L^{d-2+2\nu}
\end{equation}
on the other hand the energy gained from the external force over a
motion on a distance $u$ is
\begin{equation}
E_F = \int d^dx F u(x) \sim F L^{d+\nu}
\end{equation}
Thus in order to make the motion to the next metastable state one
needs to move a piece of the pinned system of size
\begin{equation}
L_{\rm min} \sim \left(\frac1F\right)^{\frac1{2-\nu}}
\end{equation}
The size of the minimal nucleus able to move thus grows as the
force decrease. Since the barriers to overcome grow with the size
of the object the minimum barrier to overcome ({\it assuming} that
the scaling of the barriers is {\it also} given by
(\ref{eq:scalmet}))
\begin{equation}
\Delta(F) \sim \left(\frac1F\right)^{\frac{d-2 + 2\nu}{2-\nu}}
\end{equation}
leading to a velocity
\begin{equation} \label{eq:creep}
v \propto e^{-\beta\left(\frac1F\right)^{\frac{d-2 +
2\nu}{2-\nu}}}
\end{equation}
This is a remarkable equation. It relates a dynamical property to
{\it static} exponents, and shows clearly the glassy nature of the
system. The corresponding motion has been called creep since it is
a sub-linear response. Of course the derivation given here is
phenomenological, but it was recently possible to directly derive
the creep formula from the equation of motion of the system \cite{chauve_creep_short,chauve_creep_long}.
This proved the two underlying assumptions behind the creep formula and
in particular that the scaling of the barriers and metastable
states is similar. More importantly this derivation
shows that although the formula for the velocity given by the
phenomenological derivation is correct, the actual motion is more complicated
than the naive phenomenological picture would suggest.
Indeed the phenomenological image is that a nucleus of size $L_{\rm min}$ moves through thermal
activation over a length $L_{\rm min}^\nu$, and then the process
starts again in another part of the system. In the full solution, one finds
that the motion of this small nucleus triggers an avalanche in the
system over a much larger lengthscale \cite{chauve_creep_long}. Checking for this
two scales process in simulations or actual experiments if of course a very challenging
problem.

The creep formula is quite general and will hold for interfaces as
well as periodic systems. For periodic systems, dislocations might
kill the collective behavior by providing an upper cutoff to the
size of the system that behaves collectively (as if the system was
torn into pieces). In the Bragg glass the situation is clear.
Since there are no dislocations the creep behavior persists to arbitrarily large lengthscales. Since
$\nu =0$ in the Bragg glass the creep exponent in (\ref{eq:creep})
is $\mu = 0.5$ \cite{nattermann_pinning,giamarchi_vortex_long}.
What becomes of the creep when dislocations are present is still an open
\cite{kierfeld_plastic_creep} and very challenging question.
What is sure is that one can expect a weakening of the growth of the barriers or even their saturation, when
going from the Bragg glass phase to the ``melted'' phase \cite{giamarchi_diagphas_prb}.
This is at the root of the crossing of the $I-V$ shown in Fig.~\ref{fig:peak}.
Experimental verification of the creep effects postulated above have proved difficult due
to the functional form of (\ref{eq:creep}) where the power law appears in the exponentiated
factor and requires data spanning many decades in the drive to determine the exponents
with adequate precision. Transport experiments \cite{fuchs_creep_bglass} as well as
magnetic relaxation experiments \cite{vanderbeek_relaxation_exponents} have reported
creep exponents compatible with the Bragg glass prediction, as well as weakening of the barrier growth
when going to the disordered phase. But this is clearly a very challenging and difficult
issue that would need more investigations.

\subsection{Dynamical phase diagram} \label{sec:dynamical}

Let us now turn to the problem of the nature of the moving phase.

\begin{figure}
\centerline{\includegraphics[width=\figwidth]{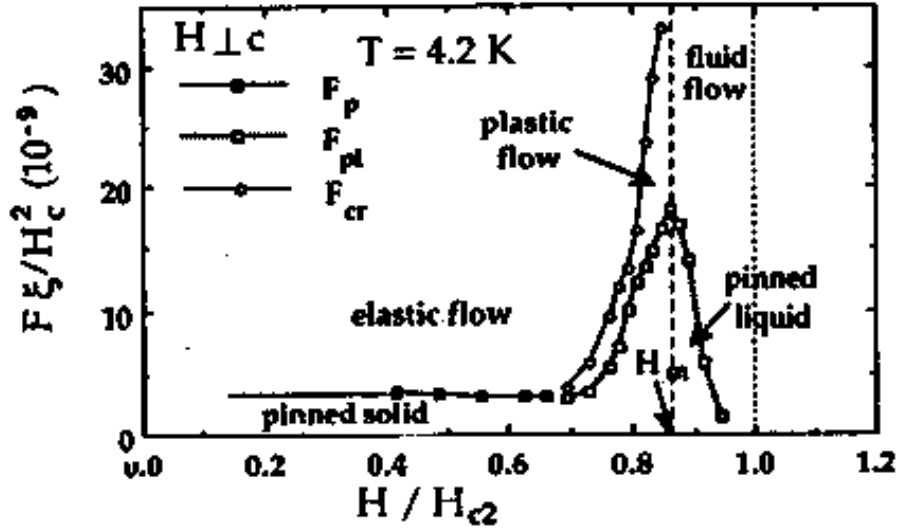}}
\caption{\label{fig:expdyndiag} A typical dynamic phase diagram in
NbSe2. Below the transition region a direct transition into a
moving ordered phase is seen while in the peak regime a pronounced
intermediate plastic flow regime is seen. At large drives above
$F_{cr}$, the moving ordered phase is established. In the
equilibrium pinned liquid phase, the crossover current is
immeasurably large.}
\end{figure}

This problem was directly prompted by experimental observations.
Indeed early measurements of the dynamics showed dramatic effects
near the peak effect (see below), led
to the construction of an experimental dynamical ''phase diagram''
\cite{bhattacharya_peak_prl} for the moving phases shown in Fig.~\ref{fig:expdyndiag}.
In addition dramatic evidence of the evolution of vortex correlations with
driving force was obtained many years ago by neutron diffraction
studies \cite{thorel_neutrons_vortex}. The Bragg peak in the pinned phase broadened
significantly at the onset of motion showing a loss of order (or
appearance of defects) and a subsequent healing of the Bragg peak
at large drives showing a reentry into an ordered moving phase.
It was thus necessary to understand the nature of the ``phases'' once the
lattice was set into motion.

\begin{figure}
\centerline{\includegraphics[width=\medfigwidth]{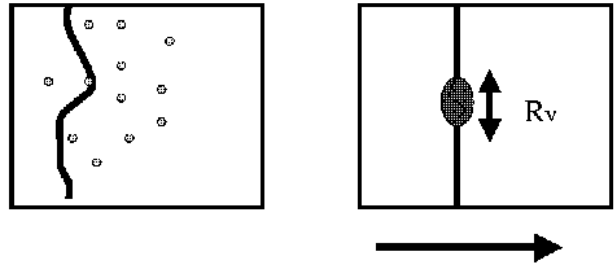}}
\caption{\label{fig:interface} When an interface is in motion, it
averages over the disorder. As a result the interface does not
feel the disorder above a certain lengthscale $R_v$ and becomes
flat again.}
\end{figure}

\begin{figure}
\centerline{\includegraphics[width=\medfigwidth]{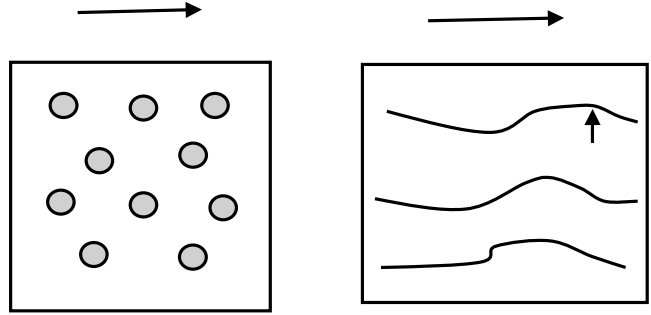}}
\caption{\label{fig:transverse} When in motion a periodic
structure averages over the Fourier components of disorder along
the direction of motion whereas the component perpendicular to
motion remains unaffected. As a result the components of the
displacements transverse to the direction of motion are the
essential ingredients to describe a moving system.}
\end{figure}

One regime in which one could think to tackle this problem is when
the lattice is moving at large velocities. Indeed in that case it
is possible to make a large velocity expansion. Such expansion was
performed with success to compute the corrections to the velocity
due to pinning \cite{larkin_largev,schmidt_hauger}, and get an
estimate of the critical current. An important step was to use the
large $v$ method to compute the displacements
\cite{koshelev_dynamics}. It was found that due to the motion the
system averages over the disorder. As a result the system do not
feel the disorder any more above a certain lengthscale and
recrystalizes. The memory of the disorder would simply be kept in
a shift of the effective temperature seen by this perfect crystal.
This picture was consistent with what was shown to be true for
interfaces, even close to threshold
\cite{nattermann_stepanow_depinning} (as shown on
Fig.~\ref{fig:interface}) and thus provided a nice explanation for
the recrystallization observed at sufficiently large velocities.
Perturbation approach along those lines has been extended in
\cite{scheidl_perturbative_phasediag}.
However the peculiarities of the periodic structure manifests themselves again,
and they does not follow this simple scenario, the way that the interfaces would.
The crucial ingredient present in periodic system is the existence
of a periodicity {\it transverse} to the direction of motion.
Because of this, the transverse components of the displacements
still feel a disorder that is non averaged by the motion. As a result
the large-$v$ expansion always breaks down, even at large velocities and
cannot be used to determine the nature of the moving phases.
To describe such moving phase the most important components are the components transverse
to the direction of motion (this is schematized in
Fig.~\ref{fig:transverse}).
The motion of
these components can be described by a quite generic equation of
motion, as explained in \cite{giamarchi_moving_prl,ledoussal_mglass_long}.
The transverse components still experience a static disorder, and as a
result the system in motion remains {\it a glass} (moving glass).

\begin{figure}
\centerline{\includegraphics[width=\smallfigwidth]{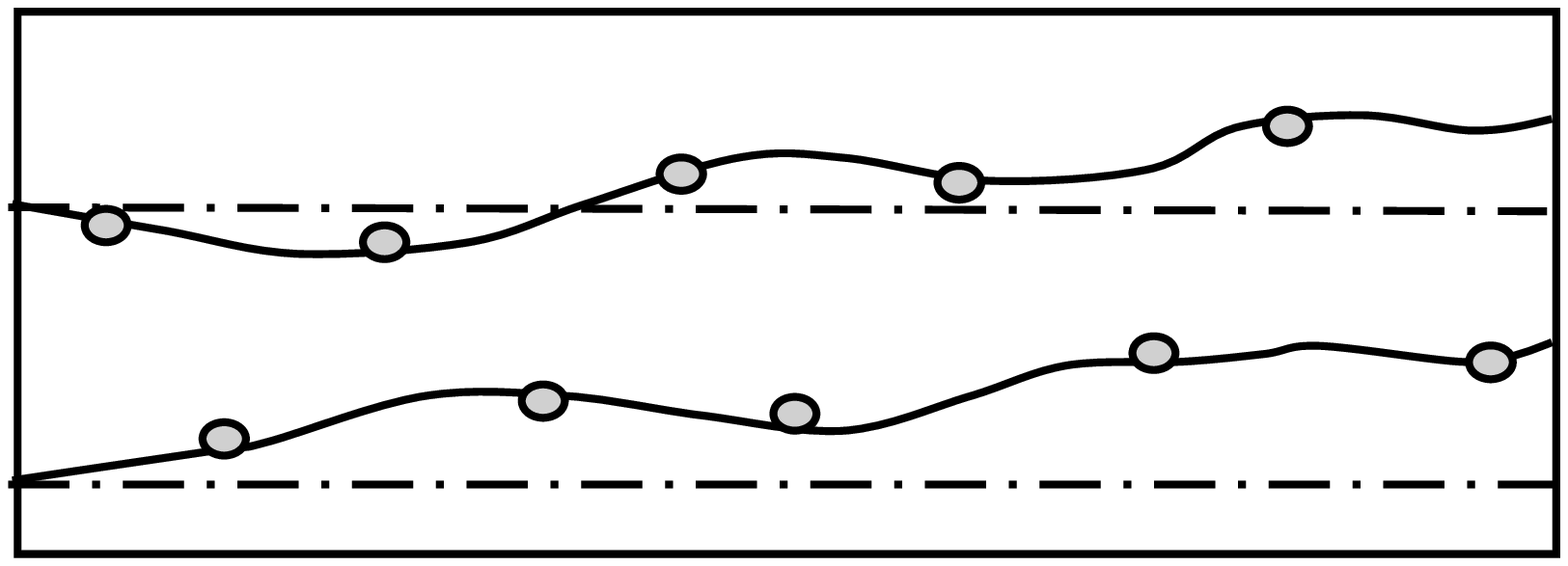}
            \includegraphics[width=\smallfigwidth]{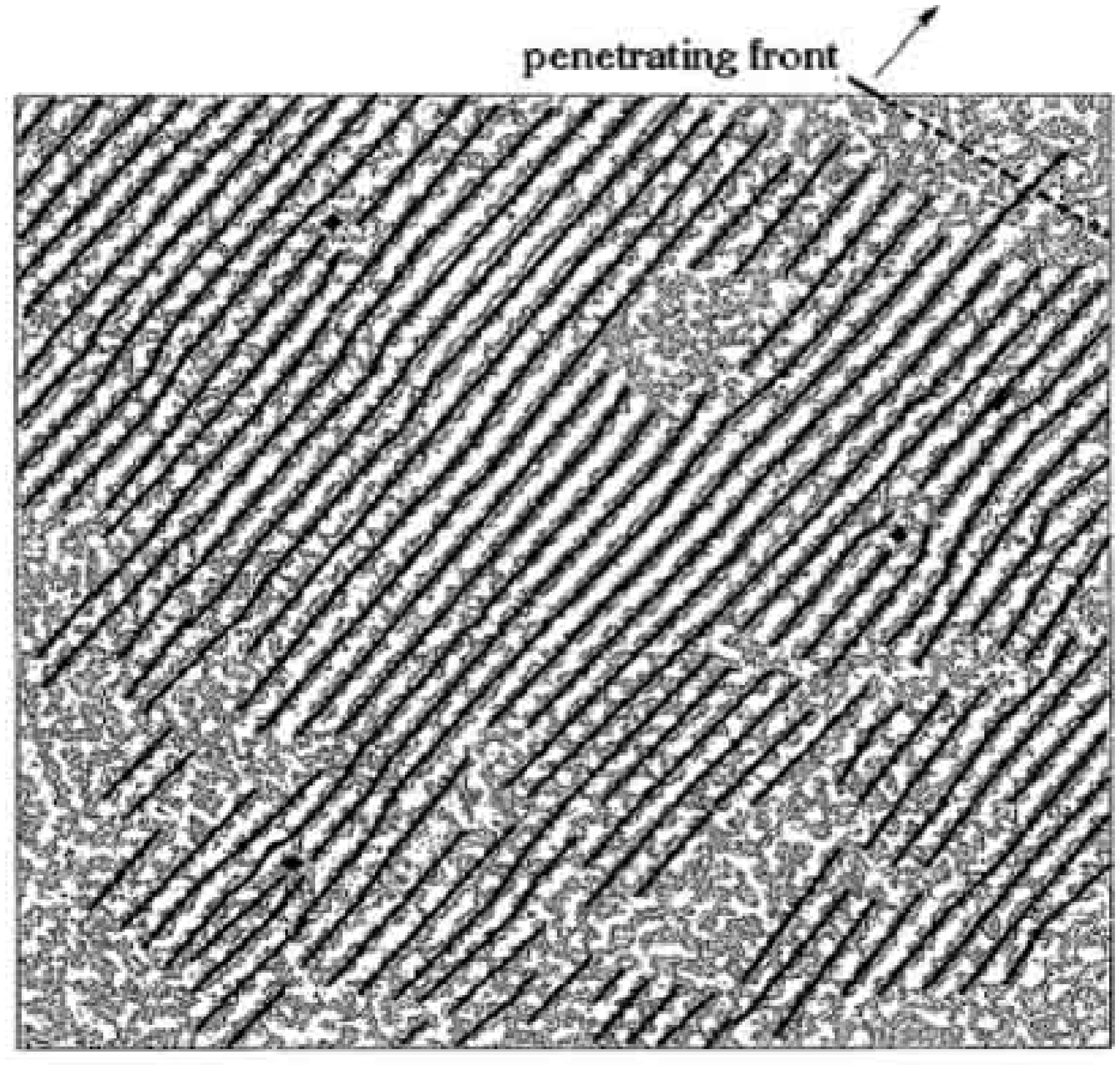}}
\caption{\label{fig:channels} In a moving glass the motion occurs
through elastic channels. The channels themselves are rough and
meander arbitrarily far from a straight line, but all particles
follows on these channels like cars on a highway. These channels have been observed in decoration
experiments, as shown on the left \cite{marchevsky_decoration_channels}}
\end{figure}

In the moving glass the motion of the particles occurs through elastic channels as shown in
Fig.~\ref{fig:channels}.
The channels are the best compromise between the elastic forces
and the static disorder still experienced by the moving system.
Like lines submitted to a static disorder the channels themselves
are rough and can meander arbitrarily far from a straight line
(displacements grow unboundedly). However although the channels
themselves are rough, the particles of the system are bound to
follow these channels and thus follow exactly the {\it same}
trajectory when in motion. Needless to say the moving system
(moving glass) is thus very different than a simple solid with a
modified temperature where the particles would just follow
straight line trajectories (with a finite thermal broadening).

When does this picture breaks down ? Clearly this should be the
result of defects appearing in the structure. For example close to
depinning, it was shown experimentally
\cite{bhattacharya_peak_prl} (see Fig.~\ref{fig:expdyndiag}) that
some of the regions of the system can remain pinned while other
parts of the system flow, leading to a plastic flow with many
defects. One thus need to check again for the stability of the
moving structure to defects. Fortunately the very existence of
channels provide a very natural framework to study the effect of
such defects: they will lead quite naturally to a coupling and a
decoupling of the channels
\cite{giamarchi_moving_prl,balents_mglass_long,ledoussal_mglass_long,%
scheidl_perturbative_phasediag}.

\begin{figure}
\centerline{\includegraphics[width=\figwidth]{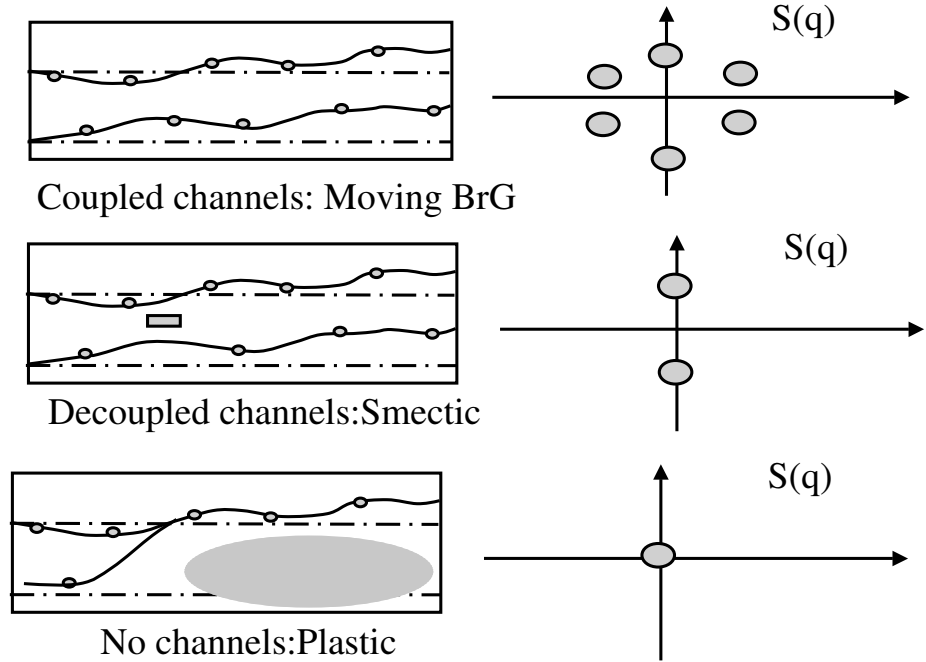}}
\caption{\label{fig:dynatrans} The various dynamical phases of a
moving periodic system.}
\end{figure}

\begin{figure}
\centerline{\includegraphics[width=\medfigwidth]{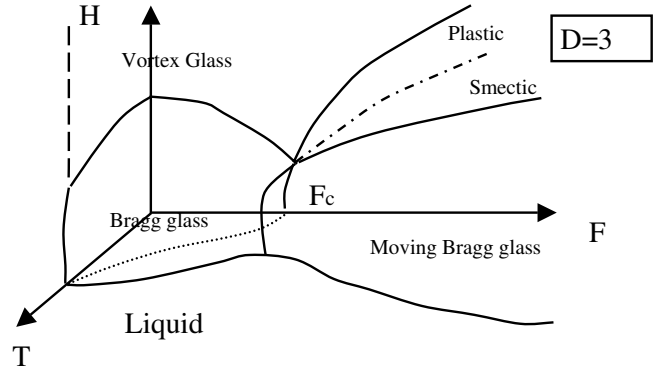}}
\caption{\label{fig:dyndiag} The dynamical phase diagram
\cite{ledoussal_mglass_long} as a function of the temperature $T$,
the magnetic field (far from $H_{c1}$) and the applied force $F$.
Both the Bragg glass and the Moving Bragg glass have perfect
topological order.}
\end{figure}

The various phases that naturally emerges in $d=3$ are thus the
ones shown in Fig.~\ref{fig:dynatrans} (a similar study can be
done for $d=2$).
At large velocity the channels are coupled and the system
possesses a perfect topological order (no defects). The moving
glass system is thus a moving Bragg glass. The structure factor
has six Bragg peaks (with algebraic powerlaw divergence) showing
that the system has quasi-long range positional order. If the
velocity is lowered a Lindemann analysis shows that defects that
appear first tend to decouple the channels. This means that
positional order along the direction of motion is lost, but since
the channel structure still exist a smectic order is preserved
(channels become then the elementary objects). As a result the
structure factor now has only two peaks. This phase is thus a
moving smectic (or a moving transverse glass, as first found in
\cite{moon_moving_numerics}).
It is important to note that in these two phases the channel
structure is preserved and described by the moving glass equation.
Both these phases are thus a moving glass. A
quite different situation can occur if the velocity is lowered
further. In that case the channel structure can be destroyed
altogether, leading to a plastic phase. Note that depending on the
amount of disorder in the system this may or may not occur, and in
$d=3$ a purely elastic depinning could be possible in principle
(in $d=2$ the depinning is always plastic). These various phases
lead to the dynamics phase diagram  shown in Fig.~\ref{fig:dyndiag}.
These various phases as well as the dynamical phase diagram have been confirmed
in numerous simulations (see e.g.
\cite{moon_moving_numerics,olson_mglass_phasediag,kolton_mglass_phases,fangohr_mglass_phasediag}).

\begin{figure}
\centerline{\includegraphics[width=\figwidth]{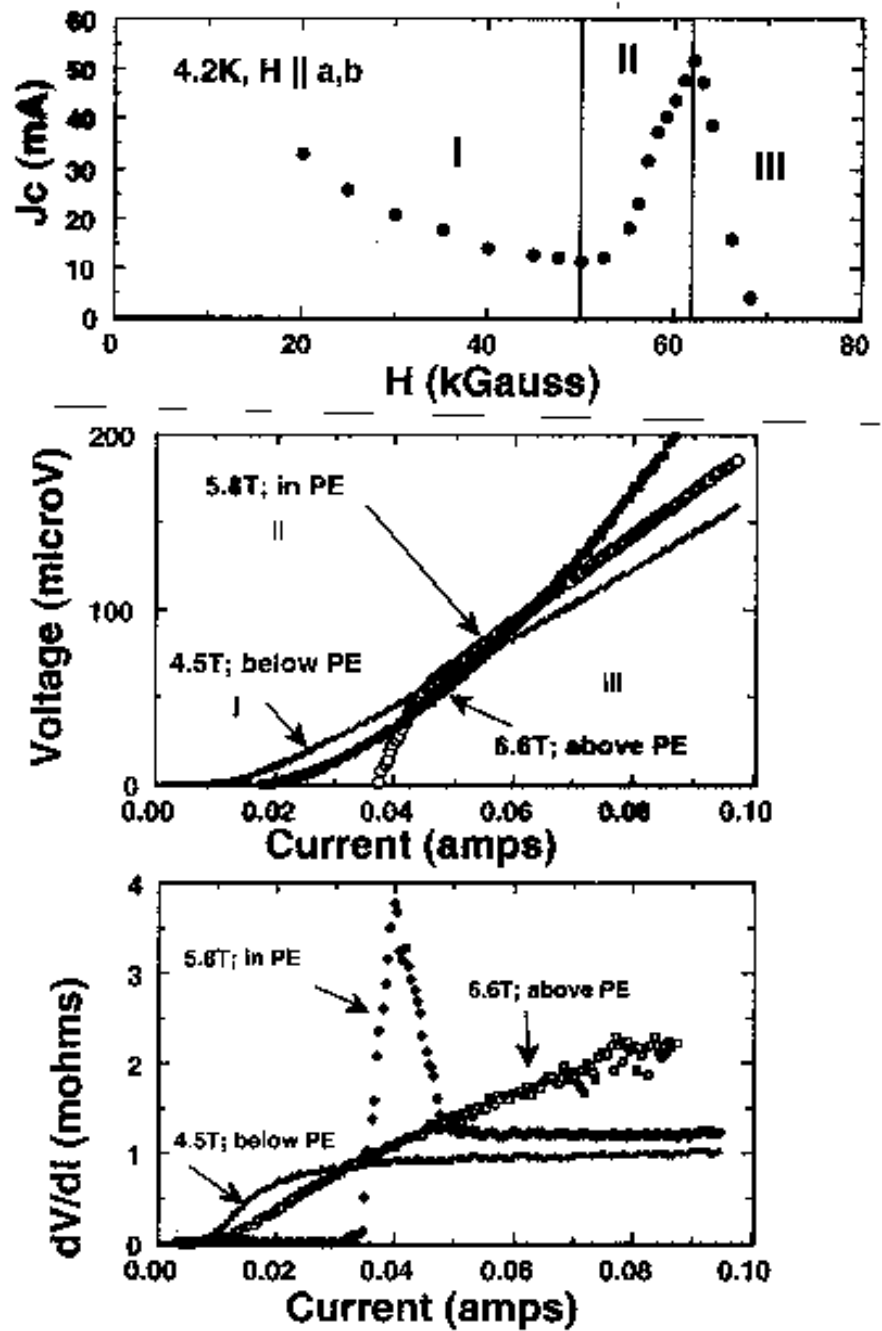}}
\caption{\label{fig:changeiv} Summary of the variation in the I-V
characteristic across the peak effect, i..e., ordered to
disordered phase transition \cite{bhattacharya_peak_prl}. The
upper panel shows the peak effect transition and marks the three
regimes of flow. The middle and lower panels show the I-V curves
and the dV/dI-I curves for the three regimes.}
\end{figure}

\begin{figure}
\centerline{\includegraphics[width=\figwidth]{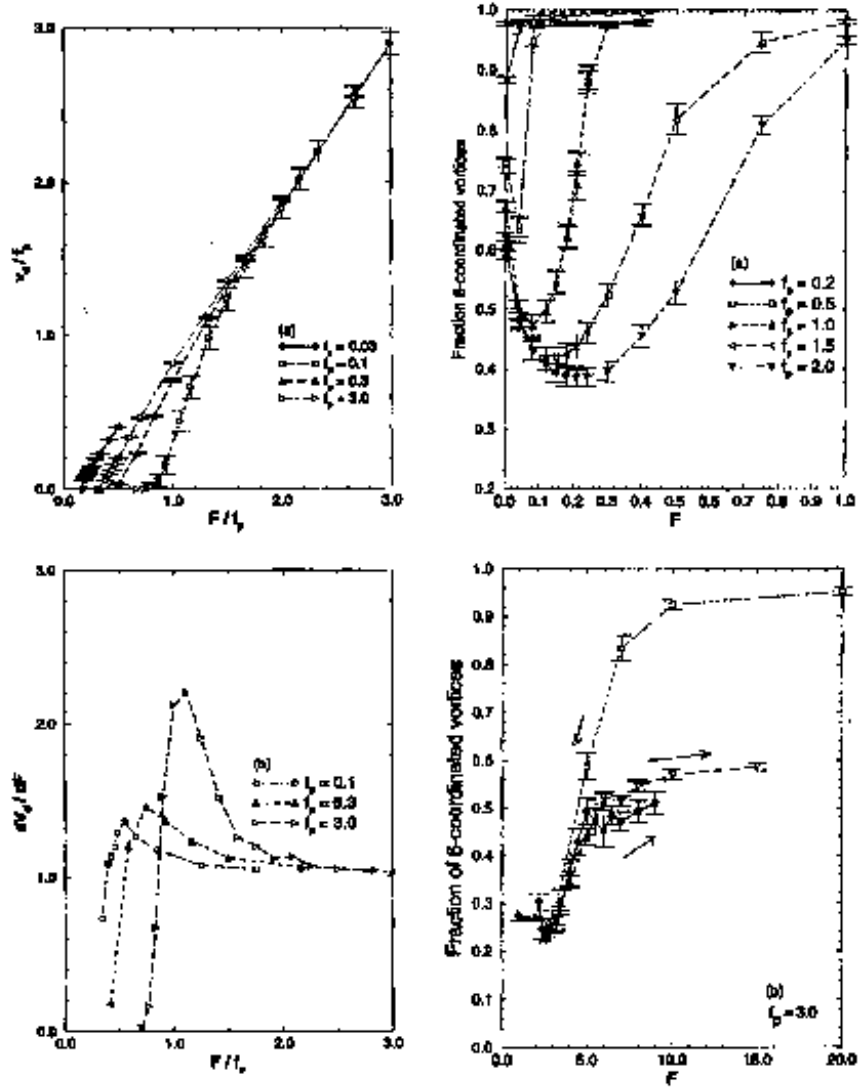}}
\caption{\label{fig:simiv} Simulation data
\cite{faleski_marchetti_dynamics} of the force-velocity curves
together with the variation of defect density with driving force.
Compare the simulation data with the experimental data for plastic
flow.}
\end{figure}

Let us now turn to the experimental analysis of the dynamics.
Early experiments on the dynamics of the vortex phases near the
peak effect \cite{bhattacharya_peak_prl} showed that not only does the critical
current rise sharply at the transition, but the $I-V$ curves also
change in character. This is shown schematically in
fig.~\ref{fig:changeiv}.
The top panel shows the peak effect at a fixed $T$
and varying $H$. Three distinct types of I-V curves are observed in
the regions marked I, II and III, shown in the middle panel. In
the peak region the curves show opposite curvature to that in the
other regions, the voltage grows convex upwards. The lower panel
illustrates this behavior through the differential resistance Rd
($=dV/dI$)for each region. In the peak region it shows a pronounced
maximum above which it rapidly decreases to a terminal asymptotic
value of the Bardeen-Stephen flux flow resistance above a
crossover current. A comparison with simulations \cite{faleski_marchetti_dynamics} shown in Fig.~\ref{fig:simiv}
suggests that the peak signifies a plastic flow region where the
vortex matter moves incoherently and a coherent flow is recovered
at high drives.
This behavior contrasts with that in region I
below the peak effect where the pinned vortex matter is in the
Bragg glass phase. In this case the flux flow resistance is
approached monotonically without a measurable signature of plastic
flow. The plastic flow regime is also accompanied with pronounced
fingerprint effect: a repeatable set of features in Rd as the
current is ramped up and down, signifying a repeatable sequence of
chunks of pinned vortex assembly depinning and joining the main
flow. This clearly establishes the defective nature of the moving
vortex matter in this regime. The contrast with this behavior in
region I is thought to represent a depinning of the Bragg glass
phase directly into a moving Bragg glass phase.

\begin{figure}
\centerline{\includegraphics[width=\figwidth]{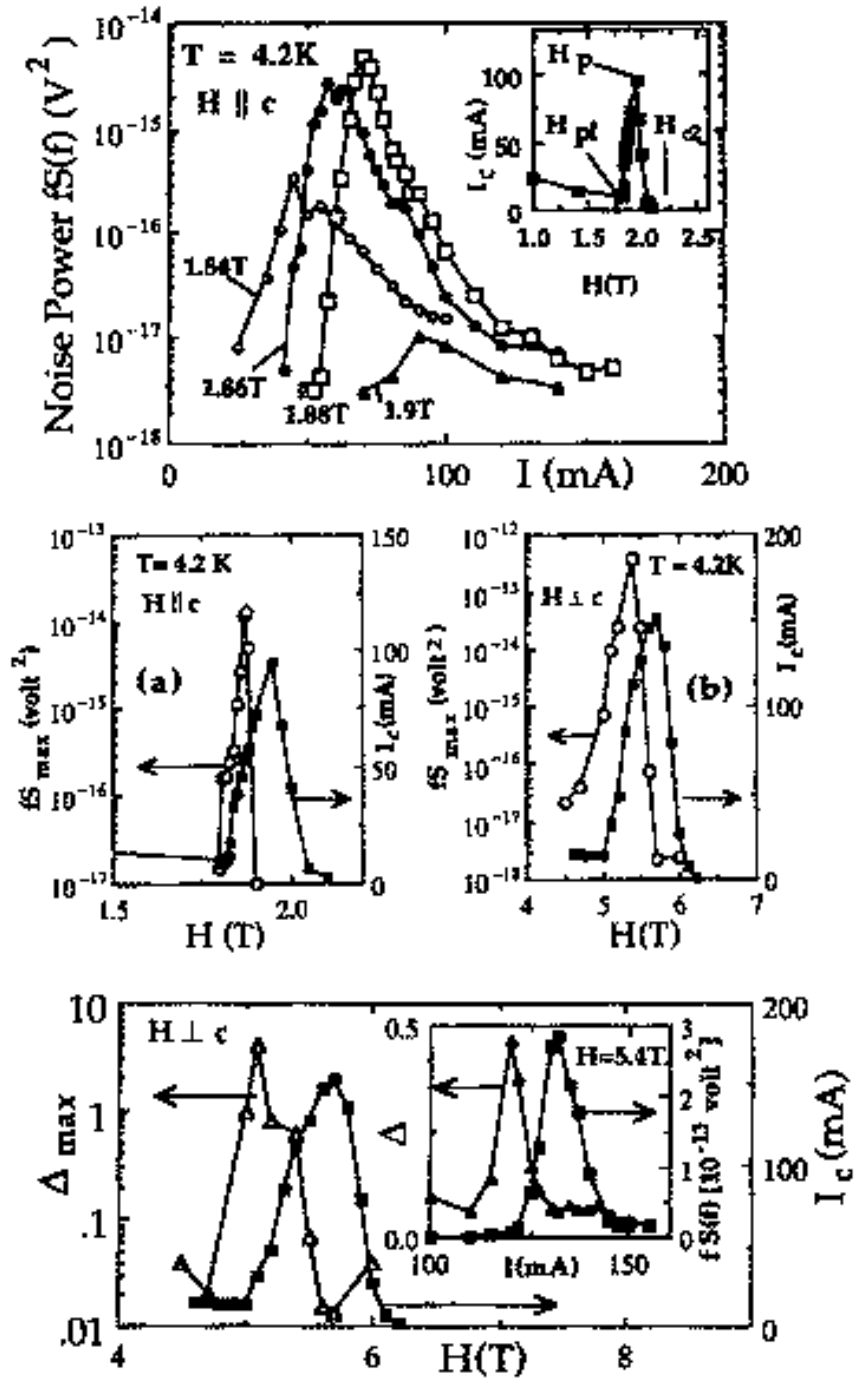}}
\caption{\label{fig:noise} Noise characteristics in the plastic
flow regime \cite{marley_broadband_noise}. The upper panel shows
the current dependence of the noise which turns on at the critical
current and depletes at large current when a moving ordered phase
is recovered. The middle panels show the field dependence of the
noise; it is restricted only to the peak regime. The lowest panel
show a measure of the non-gaussianity of the noise which is
maximum (i.e.,few noise sources) at the onset of motion supporting
the moving ``chunk'' scenario in the plastic flow regime.}
\end{figure}

A striking result is seen in the behavior of the flux flow noise
\cite{marley_broadband_noise}, summarized in Fig.~\ref{fig:noise}.
The noise is larger in the peak regime by orders of magnitude
(seen in the lower panel) and is restricted to current values
between the depinning current and the crossover current (shown in
the upper panel). The noise is, therefore, associated with an
incoherent flow of a defective moving phase and a coexistence of
moving and pinned vortex phases as was seen in detailed studies
\cite{merithew_vortex_noise,rabin_vortex_noise}. The qualitative
behavior observed in the experiment is consistent with many
simulations of the flux flow noise characteristics. However, recent
work suggests that an edge contamination
mechanism in the peak regime is responsible for triggering much of
the defective plastic flow in the bulk and is a subject under
active investigation \cite{paltiel_edge_cont,marchevsky_peakeffect_nbse}.
There have also been reports of narrow band noise \cite{togawa_mglass_noise}, i.e., noise peaks at the so-called washboard frequency. The observation
of such noise would be compatible with the moving Bragg glass phase.

More experimental evidence in
favor of the moving glass is also found in recent
decoration and STM studies of the moving vortex assembly
\cite{pardo_decoration_mglass,marchevsky_decoration_channels,troyanovski_stm_flow}.

\subsection{Metastability and history dependence}

\begin{figure}
\centerline{\includegraphics[width=\figwidth]{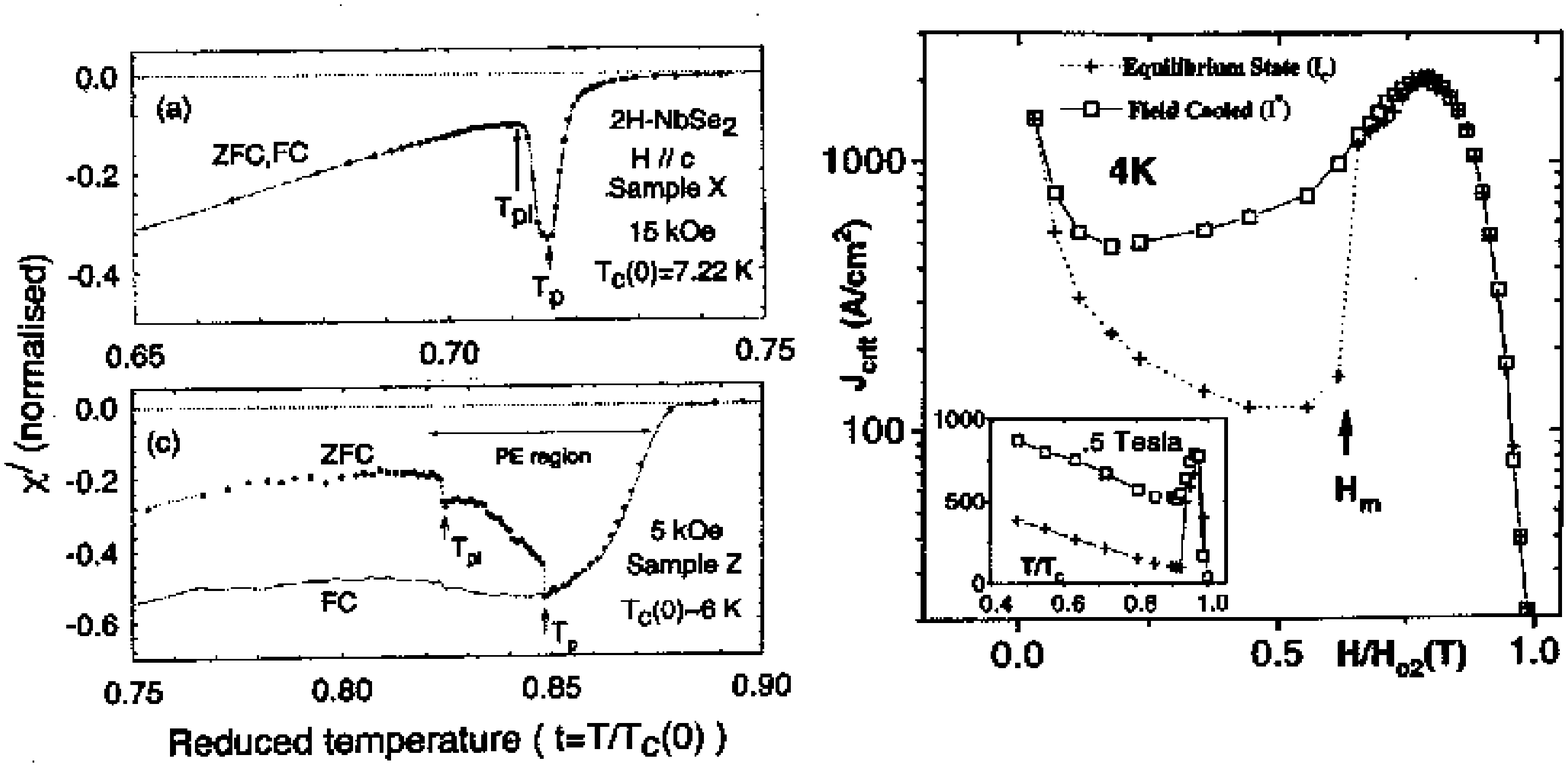}}
\caption{\label{fig:accur} History dependence of the critical
current seen with magnetic susceptibility(left)
\cite{banerjee_history_dep},  and transport(right)
\cite{henderson_history_dep}. For a very clean sample on the top
panel on the left, no history effect is seen unlike the bottom
panel for a dirty sample. The peak of the peak effect marks the
onset of an equilibrium disordered phase.}
\end{figure}

A current focus on experimental vortex phase studies is a renewed
interest in history effects that have long been known to occur in
vortex matter. In order to understand the equilibrium phase
behavior of the system, we need to ascertain that we have indeed
reached equilibrium in an experiment. However, most experiments in
low Tc systems and at low temperatures in high Tc systems as well,
show a pronounced dependence of the vortex correlations on the
thermomagnetic history of the system. A striking example is shown
in Fig.~\ref{fig:accur} where the magnetic response \cite{banerjee_history_dep,ravikumar_history_dep},
and transport critical current
\cite{henderson_history_dep} are measured for
field cooled (FC) and zero field cooled(ZFC) cases.
In the latter case one sees a pronounced peak effect but not in
the former. In other words, the FC state yields a highly
disordered vortex glass phase and the latter yields an ordered
Bragg glass phase. The question then is : which is the stable and
equilibrium state of the system and how does one find it ? A
similar question arose in the interpretation of neutron
experiments \cite{yaron_neutrons_vortex} where it was suggested
\cite{giamarchi_vortex_comment} that the observed broadening of
the lines that disappeared after a cycling above the critical
current, was due to the presence of {\it out of equilibrium}
dislocations (on the top of the equilibrium Bragg glass phase).

One possible resolution of the problem comes from a "shaking
experiment" where the FC state is subjected to a large oscillatory
magnetic field and the system evolves to the ZFC state \cite{banerjee_shake_switch}. Once in
the ZFC ordered state, below the peak effect, the system cannot be
brought to the FC disordered state regardless of any external
perturbation. It is thus reasonable to conclude that the ordered
(Bragg glass) state is the equilibrium state below the peak
effect. The FC disordered (vortex glass) phase is simply
supercooled from the liquid phase above. When pinning sets in, the
system fails to explore the phase space due to the pinning barrier
and stays frozen into glassy, disordered phase. Shaking with an ac
field then provides an annealing mechanism to bring the system in
the true ground state which is the Bragg glass. On the other hand,
shaking fails to produce an ordered state above the peak and thus
the disordered phase is indeed the ground state there. Curiously
then the ZFC state is formed by vortices entering the system at
high velocity, thereby ignoring pinning and forming a moving Bragg
glass phase from which the pinned Bragg glass phase evolves
easily. Recent aging experiments \cite{portier_age_vortex} have provided
compelling evidence in support of these conclusions. Further evidence
of a thermodynamic nature of the transition is obtained also from magnetization
anomalies from annealed vortex states \cite{ravikumar_stable_phases}.

\subsection{Edge effects}

Yet another phenomenon has begun to be explored in experimental
studies \cite{paltiel_edge_cont,marchevsky_peakeffect_nbse}. This relates to the observation that edges of a sample
provide nucleation centers for the disordered phase. The net results
are (1) the order-disorder phase transition becomes spatially
non-uniform and leads to phase coexistence that marks the width of
the peak effect region and (2) an external driving current flows
non-uniformly in the system. The contamination of the ordered
phase by the disordered phase from the edge and the subsequent annealing back to the ordered phase at larger drives
occur in a non-uniform manner leading to a variety of unusual time
and frequency dependent phenomena observed earlier.
Differentiating the effects of these processes from the bulk response
of the system, usually assumed in interpreting data and in simulations, remains a subject of current study.

\subsection{Transverse critical force}

\begin{figure}
\centerline{\includegraphics[width=\figwidth]{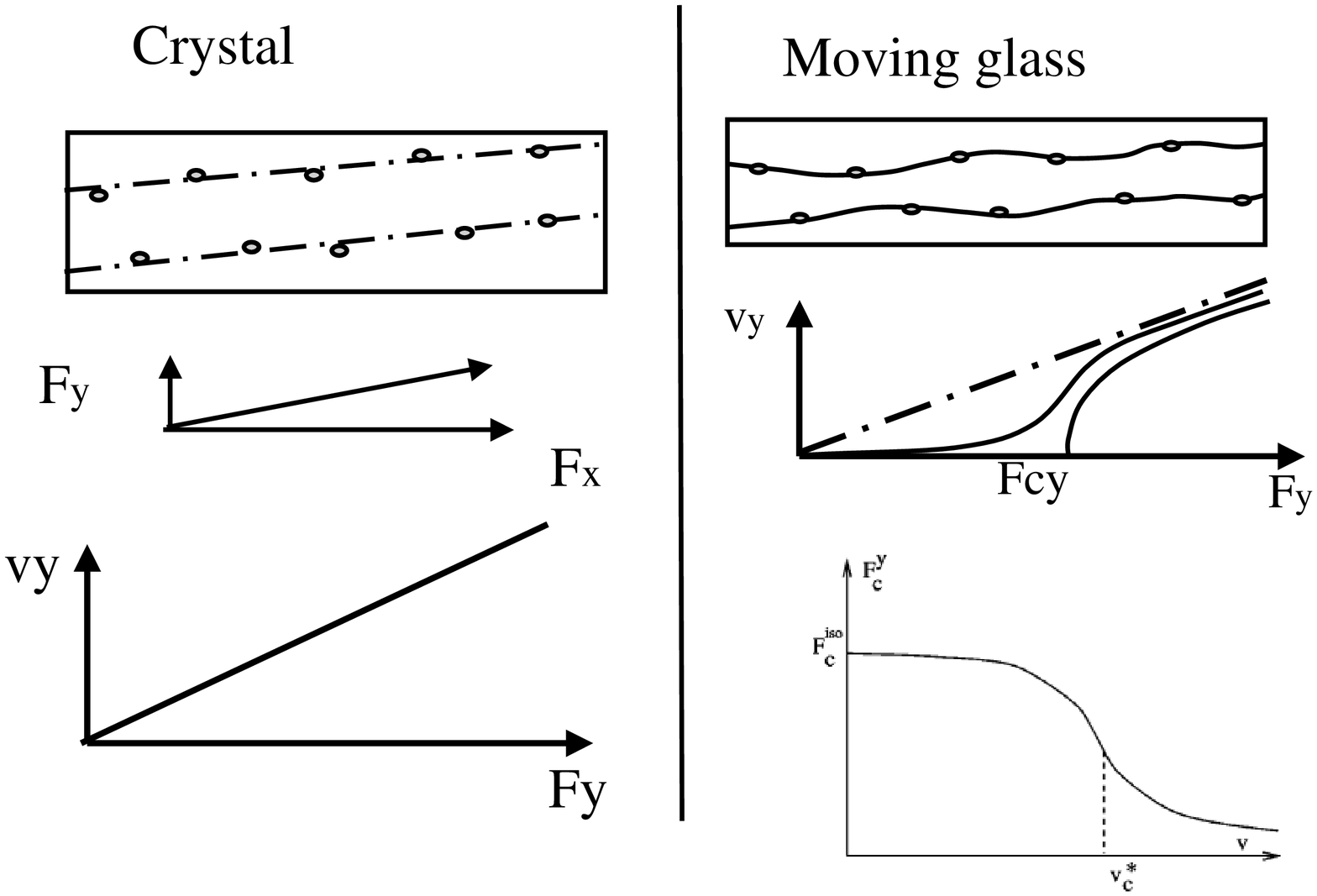}}
\caption{\label{fig:transf} Response of the system to a force
$F_y$ transverse to the direction of motion. Although a crystal
would respond linearly, the moving glass remains pinned in the
transverse direction although the particles are moving along the
direction $x$ \cite{giamarchi_moving_prl,ledoussal_mglass_long}.}
\end{figure}

One of the unexpected consequences of the presence of channels in
the moving glass phase, is the existence of transverse pinning \cite{giamarchi_moving_prl}.
Let us examine what would happen if one tried to push the moving
system sideways by applying a force $F_y$ in a direction
perpendicular to motion. This is depicted in Fig.~\ref{fig:transf}
The naive answer would be that the system is submitted to a total
force $F = F_x + F_y$ and because the modulus of this force is
larger than the threshold ($F_x > F_c$ since the system is already
sliding), the system will slide along the total force. This means
that there is a {\it linear} response $v_y$ to the applied
transverse force $F_y$. This is what would occur if the moving
system was a crystal. However in the moving glass although the
particles themselves move along the channels, the channels are
submitted to a static disorder and thus pinned. It means that if
one applies a transverse force, the channels will have to pass
barriers before they can move. As a results the system is
transversally pinned even though it is moving along the $x$
direction. The existence of this transverse critical force is a
hallmark of the moving glass. It is a fraction of the longitudinal
transverse force and decreases as the longitudinal velocity
increases as shown on Fig.~\ref{fig:transf}. The existence of such
transverse force has been confirmed in many simulations (see e.g.
\cite{moon_moving_numerics,olson_mglass_transverse,fangohr_mglass_transverse}).
Its experimental observation for vortex systems is still an experimental challenge.

\section{Conclusions and perspectives} \label{sec:conclusions}

It is thus clear from the body of work presented in these notes that the field of vortex matter has
considerably matured in the last ten years or so. Many important physical phenomena have been unravelled,
and a coherent picture starts to emerge. As far as the statics of the problem is concerned we start now
to have a good handle of the problem. The Bragg glass description has allowed to build a coherent
picture both of the phase diagram and of most of the previously poorly understood striking phenomena
such as imaging, neutrons and peak effect or more generally transport phenomena.
Important issues such as the contamination by the edge
and the necessity to untangle these effects to get the true thermodynamics properties of the system are now
understood, allowing to get reliable data. Although, understanding the dynamics is clearly much more complicated
both theoretically and experimentally, here also many progress have been made. The vortex matter has allowed
to introduce and check many crucial concepts such as the one of creep motion. For the dynamics also
it is now understood that periodicity is crucial and that strong effects of disorder can persist even
when a lattice is fast moving.

Of course the progress realized make only more apparent the many exciting issues not yet understood and that
are waiting to be solved. For the statics the nature of the high field phase above the melting field of the
Bragg glass is still to be understood. Is this phase a distinct thermodynamic phase from the liquid ? is it
a true glass in the dynamical sense ? Many difficult questions that will need the understanding of a system in which
disorder {\it and} defects (dislocations, etc.) play a crucial role. Similar question occur for the dynamics
whenever a plastic phase occurs. There is thus no doubt that understanding the role of defects and disorder
is now one of the major challenge of the field. The glassy nature of the various phases also certainly needs
more detailed investigations, in particular the ``goodies'' such as aging that have been explored in details
for systems such as spin glasses would certainly prove useful to investigate. Finally, both
theorists and experimentalists have mostly focussed for the moment on the steady state dynamics. Clearly the
issue of noise, out of equilibrium dynamics, and history dependence are challenging and crucial problems.

In addition, the material covered in these notes is only a part of the many interesting phenomena
related to the physics of vortex lattices, and more generally disordered elastic systems.
Because of space limitation, it would be impossible to cover here all these interesting aspects but
we would like to at least mention few of them.

In addition to point like impurities there is much interest, both theoretical and practical, to
introduce artificial disorder. The most popular is the one produced by heavy ion irradiation
\cite{konczykowski_columnar_first,civale_columnar_prl,vanderbeek_columnar_long},
leading to the so called columnar defects and to the Bose Glass phase \cite{nelson_columnar_long}.
Other types of disorder such as splay \cite{hwa_splay_prl} or regular pinning arrays have also been
explored. We refer the reader to the above mentioned reviews for more details on these issues.

Quite interestingly, the description of the vortices in term of
elastic objects allows to make contact with a large body of other
physical systems who share in fact the same effective physics.
This ranges from classical systems such as magnetic domain walls,
wetting contact lines, colloids, magnetic bubbles, liquid crystals
or quantum systems: charge density waves, Wigner crystals of
electrons, Luttinger liquids. These systems share the same basic
physics of the competition between elastic forces that would like
to form a nice lattice or a flat interface and disorder. This
makes the question of exploring the connections with the concepts
useful for the vortex lattices (similarities/differences)
particularly interesting and fruitful. For example, magnetic
domain walls are ideal systems to quantitatively check the creep
formula \cite{lemerle_domainwall_creep}, Wigner crystals
\cite{perruchot_transverse_wigner} or charge density waves
\cite{markovic_transverse_cdw} have been fields to test for the
existence of a transverse critical force. The question of the
observation of the noise in these various systems is also a very
puzzling question. Of course, these systems have also their own
particularities and present their own challenging problems. The
interested reader is again directed to the review papers for the
classical problems and a short review along those lines for
quantum systems can be found in \cite{giamarchi_quantum_revue}.

There is thus no doubt that the field is now growing and permeating many branches of condensed matter physics,
providing a unique laboratory to continue developing unifying concepts such as the ones born and grown for vortices
since more than half of a century. To what new developments this extraordinary richness of physical situations
will lead, only time will tell.

\section*{Acknowledgements}

We have benefitted from many invaluable discussions with many
colleagues, too numerous, to be able to thank them all here. We
would however like to especially thank G. Crabtree, P. Kes, M.
Konczykowski, W. Kwok, C. Marchetti, A. Middleton, C. Simon, K.
van der Beek, F.I.B Williams, E. Zeldov and G.T. Zimanyi. T.G.
acknowledges the many fruitful and enjoyable collaborations with
P. Le Doussal, and with P. Chauve, D. Carpentier, T. Klein, S.
Lemerle, J. Ferre. S.B. acknowledges the special contributions of
M. Higgins and thanks E. Andrei, S. Banerjee, P. deGroot, A.
Grover, M. Marchevsky, Y. Paltiel, S. Ramakrishnan, T.V.C Rao, G.
Ravikumar, M. Weissman, E. Zeldov and A. Zhukov for  fruitful
collaborations. TG would like to thank the ITP where part of these
notes were completed for hospitality and support, under the NSF
grant PHY99-07949. S.B. thanks the Tata Institute of Fundamental
Research for hospitality while materials for the lectures were
prepared.

%\bibliographystyle{prsty}
%\bibliography{totphys,cargese}

\begin{thebibliography}{100}

\bibitem{abrikosov_vortex_first}
A.~A. Abrikosov, Sov. Phys. JETP {\bf 5},  1174  (1957).

\bibitem{tinkham_book_superconductors}
M. Tinkham, {\em Introduction to Superconductivity} (Mc Graw Hill,
New York,
  1975).

\bibitem{DeGennes_supra}
P.~G. {De Gennes}, {\em Superconductivity of Metals and Alloys}
(W. A.
  Benjamin, New York, 1966).

\bibitem{blatter_vortex_review}
G. Blatter {\it et~al.}, Rev. Mod. Phys. {\bf 66},  1125  (1994).

\bibitem{brandt_review_superconductors}
H. Brandt, Rep. Prog. Phys. {\bf 58},  1465  (1995).

\bibitem{giamarchi_book_young}
T. Giamarchi and P. {Le Doussal},  in {\em Spin Glasses and Random
fields},
  edited by A.~P. Young (World Scientific, Singapore, 1998), p.\ 321,
  cond-mat/9705096.

\bibitem{nattermann_vortex_review}
T. Nattermann and S. Scheidl, Adv. Phys. {\bf 49},  607  (2000).

\bibitem{houghton_fusion_vortex}
A. Houghton, R. Pelcovits, and A. Sudbo, Phys. Rev. B {\bf 40},
6763  (1989).

\bibitem{nelson_fusion_vortex}
D. Nelson, Phys. Rev. Lett. {\bf 69},  1973  (1988).

\bibitem{charalambous_melting_rc}
M. Charalambous, J. Chaussy, and P. Lejay, Phys. Rev. B {\bf 45},
5091
  (1992).

\bibitem{safar_tricritical_prl}
H. Safar {\it et~al.}, Phys. Rev. Lett. {\bf 70},  3800  (1993).

\bibitem{safar_transport_tricritical}
H. Safar and al., Phys. Rev. B {\bf 52},  6211  (1995).

\bibitem{kwok_vortex_melting}
W. {Kwok et al.}, Phys. Rev. Lett. {\bf 72},  1088  (1994).

\bibitem{zeldov_diagphas_bisco}
E. Zeldov and Al., Nature {\bf 375},  373  (1995).

\bibitem{larkin_70}
A.~I. Larkin, Sov. Phys. JETP {\bf 31},  784  (1970).

\bibitem{fukuyama_pinning}
H. Fukuyama and P.~A. Lee, Phys. Rev. B {\bf 17},  535  (1978).

\bibitem{imry_ma}
Y. Imry and S.~K. Ma, Phys. Rev. Lett. {\bf 35},  1399  (1975).

\bibitem{larkin_ovchinnikov_pinning}
A.~I. Larkin and Y.~N. Ovchinnikov, J. Low Temp. Phys {\bf 34},
409  (1979).

\bibitem{giamarchi_vortex_short}
T. Giamarchi and P. {Le Doussal}, Phys. Rev. Lett. {\bf 72},  1530
(1994).

\bibitem{giamarchi_vortex_long}
T. Giamarchi and P. {Le Doussal}, Phys. Rev. B {\bf 52},  1242
(1995).

\bibitem{haldane_bosons}
F.~D.~M. Haldane, Phys. Rev. Lett. {\bf 47},  1840  (1981).

\bibitem{nattermann_pinning}
T. Nattermann, Phys. Rev. Lett. {\bf 64},  2454  (1990).

\bibitem{fisher_vortexglass_long}
D.~S. Fisher, M.~P.~A. Fisher, and D.~A. Huse, Phys. Rev. B {\bf
43},  130
  (1990).

\bibitem{fisher_vortexglass_short}
M.~P.~A. Fisher, Phys. Rev. Lett. {\bf 62},  1415  (1989).

\bibitem{bokil_young_vglass}
H.~S. Bokil and A.~P. Young, Phys. Rev. Lett. {\bf 74},  3021
(1995).

\bibitem{villain_cosine_realrg}
J. Villain and J.~F. Fernandez, Z. Phys. B {\bf 54},  139  (1984).

\bibitem{korshunov_variational_short}
S.~E. Korshunov, Phys. Rev. B {\bf 48},  3969  (1993).

\bibitem{emig_exponents_braggglass}
T. Emig and S. Bogner and T. Nattermann, Phys. Rev. Lett {\bf 83}
400 (1999);
  S. Bogner, T. Emig and T. Nattermann, Phys. Rev. B {\bf 63} 174501 (2001).

\bibitem{feynman_statmech}
R.~P. Feynman, {\em Statistical Mechanics} (Benjamin Reading, MA,
1972).

\bibitem{mezard_variational_global}
M. Mezard and G. Parisi, J. de Phys. I (Paris) {\bf 4}, 809
(1991); E. I.
  Shakhnovich and A. M. Gutin, J. Phys. A {\bf 22}, 1647 (1989).

\bibitem{carpentier_bglass_layered}
D. Carpentier, P. {Le Doussal}, and T. Giamarchi, Europhys. Lett.
{\bf 35},
  379  (1996).

\bibitem{kierfeld_bglass_layered}
J. Kierfeld, T. Nattermann, and T. Hwa, Phys. Rev. B {\bf 55},
626  (1997).

\bibitem{fisher_bragg_proof}
D.~S. Fisher, Phys. Rev. Lett. {\bf 78},  1964  (1997).

\bibitem{gingras_dislocations_numerics}
M.~J.~P. Gingras and D.~A. Huse, Phys. Rev. B {\bf 53},  15193
(1996).

\bibitem{vanotterlo_bragg_numerics}
A.~V. Otterlo, R. Scalettar, and G. Zimanyi, Phys. Rev. Lett. {\bf
81},  1497
  (1998).

\bibitem{marchevsky_thesis}
M. Marchevsky, Ph.D. thesis, University of Leiden, 1998.

\bibitem{hess_stm_vortex}
H. {Hess et al.}, Phys. Rev. Lett. {\bf 62},  214  (1989).

\bibitem{harada_lorentz_vortex}
K. {Harada et al.}, Nature {\bf 360},  51  (1992).

\bibitem{ling_neutrons_bragg}
X. {Ling et al.}, Phys. Rev. Lett. {\bf 86},  126  (2001).

\bibitem{grier_decoration_manips}
D.~G. Grier {\it et~al.}, Phys. Rev. Lett. {\bf 66},  2270
(1991).

\bibitem{kim_decorations_nbse}
P. Kim, Z. Yao, and C.~A. Bolle, Phys. Rev. B {\bf 60},  R12589
(1999).

\bibitem{klein_brglass_nature}
T. {Klein et al.}, Nature {\bf 413},  404  (2001).

\bibitem{khaykovich_diagphas_bisco}
B. Khaykovich and al., Phys. Rev. Lett. {\bf 76},  2555  (1996).

\bibitem{cubbit_neutrons_bscco}
R. Cubbit and {al.}, Nature {\bf 365},  407  (1993).

\bibitem{gammel_neutrons_Nb}
P.~L. Gammel {\it et~al.}, Phys. Rev. Lett. {\bf 80},  833
(1998).

\bibitem{ertas_diagphas_bisco}
D. Ertas and D.~R. Nelson, Physica C {\bf 272},  79  (1996).

\bibitem{giamarchi_diagphas_prb}
T. Giamarchi and P. {Le Doussal}, Phys. Rev. B {\bf 55},  6577
(1997).

\bibitem{kierfeld_diagphas_bisco}
J. Kierfeld, Physica C {\bf 300},  171  (1998).

\bibitem{koshelev_diagphas_bisco}
A.~E. Koshelev and V.~M. Vinokur, Phys. Rev. B {\bf 57},  8026
(1998).

\bibitem{bouquet_melting_ybco}
F. {Bouquet et al.}, Nature {\bf 411},  448  (2001).

\bibitem{welp_magnetization_jump}
U. {Welp et al.}, Phys. Rev. Lett. {\bf 76},  4908  (1996).

\bibitem{schilling_heat_vortex}
A. Schilling, R.~A. Fisher, and G.~W. Crabtree, Nature {\bf 382},
791  (1996).

\bibitem{deligiannis_diagphas_ybco}
K. Deligiannis and al., Phys. Rev. Lett. {\bf 79},  2121  (1997).

\bibitem{nishizaki_diagphas_ybco}
T. Nishizaki and {\it et al.}, Phys. Rev. B {\bf 61},  3649
(2000).

\bibitem{kokkaliaris_diagphas_ybco}
S. Kokkaliaris, A.~A. Zhukov, and P.~A.~J. {de Groot}, Phys. Rev.
B {\bf 61},
  3655  (2000).

\bibitem{wordenweber_kes_peak}
R. Wordenweber, P. Kes, and C. Tsuei, Phys. Rev. B {\bf 33},  3172
(1986).

\bibitem{bhattacharya_peak_prl}
S. Bhattacharya and M.~J. Higgins, Phys. Rev. Lett. {\bf 70},
2617  (1993).

\bibitem{higgins_second_peak}
M.~J. Higgins and S. Bhattacharya, Physica C {\bf 257},  232
(1996).

\bibitem{rao_musr_peak}
T.~V.~C. {Rao et al.}, Physica C {\bf 299},  267  (1998).

\bibitem{ghosh_reentrant_peak}
K. {Ghosh et al.}, Phys. Rev. Lett. {\bf 76},  4600  (1996).

\bibitem{banerjee_reentrant_peak}
S.~S. {Banerjee et al.}, Europhys. Lett. {\bf 44},  91  (1998).

\bibitem{aegerter_YBCO_musr}
C.~M. {Aegerter et al.}, Phys. Rev. B {\bf 57},  1253  (1998).

\bibitem{ishida_peak_melting}
T. {Ishida et al.}, Phys. Rev. B {\bf 56},  5128  (1997).

\bibitem{menon_phasediag_univ}
G.~I. Menon, 2001, cond-mat/0103013.

\bibitem{vanotterloo_IV_peakeffect}
A. {van Otterloo et al.}, Phys. Rev. Lett. {\bf 84},  2493
(2000).

\bibitem{chauve_creep_long}
P. Chauve, T. Giamarchi, and P. {Le Doussal}, Phys. Rev. B {\bf
62},  6241
  (2000).

\bibitem{anderson_kim}
P.~W. Anderson and Y.~B. Kim, Rev. Mod. Phys. {\bf 36},  39
(1964).

\bibitem{nattermann_rfield_rbond}
T. Nattermann, Europhys. Lett. {\bf 4},  1241  (1987).

\bibitem{ioffe_creep}
L.~B. Ioffe and V.~M. Vinokur, J. Phys. C {\bf 20},  6149  (1987).

\bibitem{feigelman_collective}
M. Feigelman, V.~B. Geshkenbein, A.~I. Larkin, and V. Vinokur,
Phys. Rev. Lett.
  {\bf 63},  2303  (1989).

\bibitem{chauve_creep_short}
P. Chauve, T. Giamarchi, and P. {Le Doussal}, Europhys. Lett. {\bf
44},  110
  (1998).

\bibitem{kierfeld_plastic_creep}
J. Kierfeld, H. Nordborg, and V.~M. Vinokur, Phys. Rev. Lett. {\bf
85},  4948
  (2000).

\bibitem{fuchs_creep_bglass}
D. Fuchs and al., Phys. Rev. Lett. {\bf 80},  4971  (1998).

\bibitem{vanderbeek_relaxation_exponents}
C.~J. {van der Beek et al.}, Physica C {\bf 341},  1279  (2000).

\bibitem{thorel_neutrons_vortex}
R. Thorel and Al., J. Phys. (Paris) {\bf 34},  447  (1973).

\bibitem{larkin_largev}
A.~I. Larkin and Y.~N. Ovchinnikov, Sov. Phys. JETP {\bf 38},  854
(1974).

\bibitem{schmidt_hauger}
A. Schmidt and W. Hauger, J. Low Temp. Phys {\bf 11},  667
(1973).

\bibitem{koshelev_dynamics}
A.~E. Koshelev and V.~M. Vinokur, Phys. Rev. Lett. {\bf 73},  3580
(1994).

\bibitem{nattermann_stepanow_depinning}
T. Nattermann, S. Stepanow, L.~H. Tang, and H. Leschhorn, J. Phys.
(Paris) {\bf
  2},  1483  (1992).

\bibitem{scheidl_perturbative_phasediag}
S. Scheidl and V. Vinokur, Phys. Rev. B {\bf 57},  13800  (1998).

\bibitem{giamarchi_moving_prl}
T. Giamarchi and P. {Le Doussal}, Phys. Rev. Lett. {\bf 76},  3408
(1996).

\bibitem{ledoussal_mglass_long}
P. {Le Doussal} and T. Giamarchi, Phys. Rev. B {\bf 57},  11356
(1998).

\bibitem{marchevsky_decoration_channels}
M. Marchevsky and al., Phys. Rev. Lett. {\bf 78},  531  (1997).

\bibitem{balents_mglass_long}
L. Balents, C. Marchetti, and L. Radzihovsky, Phys. Rev. B {\bf
57},  7705
  (1998).

\bibitem{moon_moving_numerics}
K. Moon and al., Phys. Rev. Lett. {\bf 77},  2378  (1997).

\bibitem{olson_mglass_phasediag}
C.~J. {Olson et al.}, Phys. Rev. Lett. {\bf 81},  3757  (1998).

\bibitem{kolton_mglass_phases}
A. {Kolton et al.}, Phys. Rev. Lett. {\bf 83},  3061  (1999).

\bibitem{fangohr_mglass_phasediag}
H. Fangohr, S.~J. Cox, and P.~A.~J. {de Groot}, Phys. Rev. B {\bf
64},  064505
  (2001).

\bibitem{faleski_marchetti_dynamics}
A.~A.~M. M.~C.~Faleski, M. C.~Marchetti, Phys. Rev. B {\bf 54},
12427  (1996).

\bibitem{marley_broadband_noise}
A.~C. Marley, M.~J. Higgins, and S. Bhattacharya, Phys. Rev. Lett.
{\bf 74},
  3029  (1995).

\bibitem{merithew_vortex_noise}
R.~D. {Merithew et al.}, Phys. Rev. Lett. {\bf 77},  3197  (1996).

\bibitem{rabin_vortex_noise}
M. {Rabin et al.}, Phys. Rev. B {\bf 57},  R720  (1998).

\bibitem{paltiel_edge_cont}
Y. {Paltiel et al.}, Nature {\bf 403},  398  (2000).

\bibitem{marchevsky_peakeffect_nbse}
M. Marchevsky, M.~J. Higgins, and S. Bhattacharya, Nature {\bf
409},  591
  (2001).

\bibitem{togawa_mglass_noise}
Y. {Togawa et al.}, Phys. Rev. Lett. {\bf 85},  3716  (2000).

\bibitem{pardo_decoration_mglass}
F. Pardo {\it et~al.}, Nature {\bf 396},  348  (1998).

\bibitem{troyanovski_stm_flow}
A.~M. {Troyanovski et al.}, Nature {\bf 399},  665  (1999).

\bibitem{banerjee_history_dep}
S.~S. {Banerjee et al.}, Phys. Rev. B {\bf 58},  995  (1998).

\bibitem{ravikumar_history_dep}
G. {Ravikumar et al.}, Phys. Rev. B {\bf 57},  R11069  (1998).

\bibitem{henderson_history_dep}
W. {Henderson et al.}, Phys. Rev. Lett. {\bf 77},  2077  (1996).

\bibitem{yaron_neutrons_vortex}
U. Yaron and al., Phys. Rev. Lett. {\bf 73},  2748  (1994).

\bibitem{giamarchi_vortex_comment}
T. Giamarchi and P. {Le Doussal}, Phys. Rev. Lett. {\bf 75},  3372
(1995).

\bibitem{banerjee_shake_switch}
S.~S. {Banerjee et al.}, Appl. Phys. Lett. {\bf 74},  126  (1999).

\bibitem{portier_age_vortex}
F. {Portier et al.}, 2001, cond-mat 0109077.

\bibitem{ravikumar_stable_phases}
G. {Ravikumar et al.}, Phys. Rev. B {\bf 63},  0240505  (2001).

\bibitem{olson_mglass_transverse}
C.~J. {Olson et al.}, Phys. Rev. B {\bf 61},  R3811  (1999).

\bibitem{fangohr_mglass_transverse}
H. Fangohr, P.~A.~J. {de Groot}, and S.~J. Cox, Phys. Rev. B {\bf
63},  064501
  (2001).

\bibitem{konczykowski_columnar_first}
M. Konczykowski and al., Phys. Rev. B {\bf 44},  7167  (1991).

\bibitem{civale_columnar_prl}
L. Civale and al., Phys. Rev. Lett. {\bf 67},  648  (1991).

\bibitem{vanderbeek_columnar_long}
C.~J. {van der Beek} and al., Phys. Rev. B {\bf 51},  15492
(1995).

\bibitem{nelson_columnar_long}
D.~R. Nelson and V.~M. Vinokur, Phys. Rev. B {\bf 48},  13060
(1993).

\bibitem{hwa_splay_prl}
T. Hwa, P. {Le Doussal}, D.~R. Nelson, and V.~M. Vinokur, Phys.
Rev. Lett. {\bf
  71},  3545  (1993).

\bibitem{lemerle_domainwall_creep}
S. Lemerle and {et al.}, Phys. Rev. Lett. {\bf 80},  849  (1998).

\bibitem{perruchot_transverse_wigner}
F. {Perruchot et al.}, Physica B {\bf 256},  587  (1998).

\bibitem{markovic_transverse_cdw}
N. Markovic, N. Dohmen, and H. {van der Zant}, Phys. Rev. Lett.
{\bf 84},  534
  (2000).

\bibitem{giamarchi_quantum_revue}
T. Giamarchi and E. Orignac,  in {\em New Theoretical Approaches
to Strongly
  Correlated Systems}, Vol.~23 of {\em NATO SCIENCE SERIES: II: Mathematics,
  Physics and Chemistry}, edited by A.~M. Tsvelik (Kluwer Academic Publishers,
  Dordrecht, 2001), cond-mat/0005220.

\end{thebibliography}

\end{document}